\documentclass[twocolumn]{aastex62}

\submitjournal{ApJ}

\shortauthors{Beane et al.}
\shorttitle{Galactic Dynamical Analysis and Midplane Fluctuations}

\usepackage{graphicx}
\usepackage{gensymb}
\usepackage{bm}
\usepackage{mathtools}
\usepackage{makecell}
\usepackage{scrextend}

\newcommand{\Msun}{\ensuremath{\text{M}_\odot}}
\newcommand{\pc}{\text{pc}}
\newcommand{\kpc}{\text{kpc}}
\newcommand{\Myr}{\text{Myr}}
\newcommand{\Gyr}{\text{Gyr}}
\newcommand{\kms}{\text{km}\,\text{s}^{-1}}
\newcommand{\actunit}{\text{kpc}\,\kms}

\newcommand{\mi}{\texttt{m12i}}
\newcommand{\mf}{\texttt{m12f}}
\newcommand{\mm}{\texttt{m12m}}

\newcommand{\sgra}{Sgr~A\textsuperscript{*}}

\newcommand{\abs}[1]{\left| #1 \right|}
\newcommand{\avg}[1]{\left< #1 \right>}
\newcommand{\z}{z_r}

\newcommand{\beq}{\begin{equation}}
\newcommand{\eeq}{\end{equation}}

\newcommand{\thin}{\texttt{thin-disk}}
\newcommand{\thick}{\texttt{thick-disk}}
\newcommand{\halo}{\texttt{halo}}

\newcommand{\cca}{Center for Computational Astrophysics, Flatiron Institute,
162 5th Ave., New York, NY 10010, USA}
\newcommand{\penn}{Department of Physics \& Astronomy, University of
Pennsylvania, 209 South 33rd St., Philadelphia, PA 19104, USA}
\newcommand{\amnh}{Department of Astrophysics, American Museum of Natural
History, Central Park West at 79th St., New York, NY 10024, USA}
\newcommand{\columbia}{Department of Astronomy, Columbia University, 550 West
120th St., New York, NY 10027, USA}
\newcommand{\victoria}{Department of Physics \& Astronomy, University of
Victoria, 3800 Finnerty Rd., Victoria, BC, V8P 4HN, Canada}
\newcommand{\nyuphys}{Center for Cosmology and Particle Physics, Department of
Physics, New York University, 726 Broadway, New York, NY 10003, USA}
\newcommand{\nyucds}{Center for Data Science, New York University, 60 5th
Ave., New York, NY 10011, USA}
\newcommand{\mpia}{Max-Planck-Institut f\"ur Astronomie, K\"onigstuhl 17,
69117 Heidelberg, Germany}

\begin{document}

\title{The Implications of Local Fluctuations in the Galactic Midplane for
Dynamical Analysis in the \textit{Gaia} Era}

\correspondingauthor{Angus Beane}
\email{abeane@sas.upenn.edu}

\author[0000-0002-8658-1453]{Angus Beane}
\affil{\cca}
\affil{\penn}

\author[0000-0003-3939-3297]{Robyn E. Sanderson}
\affil{\penn}
\affil{\cca}

\author{Melissa K. Ness}
\affil{\columbia}
\affil{\cca}

\author{Kathryn V. Johnston}
\affil{\columbia}
\affil{\cca}

\author[0000-0001-8275-9181]{Douglas Grion Filho}
\affil{\columbia}

\author[0000-0003-0064-4060]{Mordecai-Mark Mac Low}
\affil{\amnh}
\affil{\cca}

\author{Daniel Angl\'es-Alc\'azar}
\affil{\cca}

\author[0000-0003-2866-9403]{David W. Hogg}
\affil{\nyuphys}
\affil{\nyucds}
\affil{\cca}
\affil{\mpia}

\author{Chervin F. P. Laporte}
\altaffiliation{CITA National Fellow}
\affil{\victoria}

\begin{abstract}

Orbital properties of stars, computed from their six-dimensional phase space
measurements and an assumed Galactic potential, are used to understand the
structure and evolution of the Galaxy. Stellar actions, computed from orbits,
have the attractive quality of being invariant under certain assumptions and
are therefore used as quantitative labels of a star's orbit. We report a
subtle but important systematic error that is induced in the actions as a
consequence of local midplane variations expected for the Milky Way. This
error is difficult to model because it is non-Gaussian and bimodal, with
neither mode peaking on the null value. An offset in the vertical position of
the Galactic midplane of $\sim15\,\pc$ for a thin disk-like orbit or $\sim
120\,\pc$ for a thick disk-like orbit induces a $25\%$ systematic error in the
vertical action $J_z$. In FIRE simulations of Milky Way-mass galaxies, these
variations are on the order of $\sim100\,\pc$ at the solar circle. From
observations of the mean vertical velocity variation of
$\sim5\textup{--}10\,\kms$ with radius, we estimate that the Milky Way
midplane variations are $\sim60\textup{--}170\,\pc$, consistent with
three-dimensional dust maps. Action calculations and orbit integrations, which
assume the global and local midplanes are identical, are likely to include
this induced error, depending on the volume considered. Variation in the local
standard of rest or distance to the Galactic center causes similar issues. The
variation of the midplane must be taken into account when performing dynamical
analysis across the large regions of the disk accessible to \textit{Gaia} and
future missions.

\end{abstract}


\section{Introduction} \label{sec:intro} 
Our understanding of the Milky Way is currently undergoing a revolution as a
result of \textit{Gaia} Data Release 2 (DR2). Recent major discoveries include
the affirmation of remnants of a major merger \citep{2018ApJ...860L..11K,
2018MNRAS.478..611B, 2018Natur.563...85H, 2019MNRAS.486..378L,
2019MNRAS.482.3426M} hinted at in pre-\textit{Gaia} work
\citep[e.g.,][]{2005MNRAS.359...93M, 2011MNRAS.412.1203N}, a phase-space
``spiral'' in the solar neighborhood \citep{2018Natur.561..360A} possibly
indicating local substructure infall \citep{2018MNRAS.481.1501B,
2019MNRAS.485.3134L}, and a gap suggestive of perturbation by a dark matter
subtructure in the tidal stream GD1 \citep{2018ApJ...863L..20P,
2018arXiv181103631B}. These discoveries all indicate that the Milky Way's
stellar distribution, which demonstrably departs significantly from
axisymmetry, is undergoing phase mixing and dynamical interactions across a
range of spatial and temporal scales.

The assumption of a global, axisymmetric Galactocentric coordinate system
\citep{2008gady.book.....B} underlies much of the quantitative analysis of the
mechanisms that give rise to these signatures. In order to construct such a
system, the Sun's relative position and velocity must be defined and measured
both precisely and accurately. This involves determining the angular position
of and distance to the Galactic center, the orientation of and distance to the
Galactic midplane, and the local standard of rest (LSR). We review and discuss
the observational efforts to measure these parameters in
Section~\ref{ssec:coord_off}.

Once a Galactocentric coordinate system has been established and a
six-dimensional (6D) phase space measurement of a star has been made, it is
often desirable to convert this measurement into action space to concisely
summarize its projected orbit, model the stellar distribution function, or
find stars with similar dynamical properties. Actions are conserved quantities
that describe the orbit of a star under the assumption of regular, bound
orbits in a system where the equations of motion are separable in a particular
coordinate system. They are the cyclical integral of the canonical momentum
over its conjugate position:
\beq\label{eq:actions} 
J_i \equiv \frac{1}{2\pi}
\oint_{\text{orbit}}p_i\,\text{d}x_i\text{,}
\eeq 
where $p_i$ are the conjugate momenta. Under the assumption of axisymmetry,
$i=R,\phi,z$ are the radial, azimuthal, and vertical coordinates respectively
in a cylindrical coordinate system. In a slowly-evolving axisymmetric
potential, these actions are invariant and $J_{\phi} \equiv L_z$, where $L_z$
is the $z$-component of the angular momentum
\citep{2008gady.book.....B,2014RvMP...86....1S}.

With the advent of 6D phase-space measurements over a relatively large
($\gtrsim 2$ kpc) volume from the \textit{Gaia} satellite, the study of
stellar actions has gained new popularity. One reason is dimensionality
reduction\,---\,an individual stellar orbit is concisely described by three
actions, as opposed to six phase space coordinates. Second, under the
assumption of a phase-mixed system, the dynamical properties of a population
of stars should be uniquely a function of their actions and independent of the
conjugate angles. This allows one to use actions to study the relationship
between \emph{orbital} properties of stars and other intrinsic, and, at least
partially, invariant properties such as age or metallicity
\citep{2018ApJ...867...31B, 2018arXiv180803278T, 2018MNRAS.481.4093S,
2019arXiv190304030G, 2019arXiv190309320D, 2019MNRAS.486.1167B}. Actions also
provide a convenient basis for constructing models of the stellar distribution
function \citep[e.g.,][]{1915MNRAS..76...70J, 1985ApJ...295..388V,
2017ApJ...839...61T}, or for associating stars with similar dynamical
properties, e.g., to potentially determine membership in moving groups.

If the system being considered departs from axisymmetry in a significant
and/or non-adiabatic way, the actions computed using an axisymmetric
approximation to the true potential can exhibit cyclic dependence on the
orbital phase (or time at which the star's position and velocity are
observed), large-scale migration, or diffusion from their initial values. In
the Milky Way, stellar actions are expected to diffuse on short time scales
due to scattering with gas clouds and to evolve on longer time scales in the
case of orbits near resonances with spiral arms, bar(s), and other large scale
perturbations \citep{2014RvMP...86....1S}. For this reason, actions have been
used to study stellar scattering in the Milky Way disk using the improved
astrometry of \textit{Gaia} DR2 and various age catalogues
\citep{2018ApJ...867...31B, 2018arXiv180803278T}. Actions have also been used
to study different models of spiral structure in the Milky Way
\citep{2019MNRAS.tmp..155S}. Characteristics of the distribution of stars in
the extended solar neighborhood in action space are discussed in
\citet{2019MNRAS.484.3291T}.

The true Galactic potential is not strictly axisymmetric, beyond even
the typically-quoted gas clouds, spiral arms, and bar(s). The presence of disk
oscillations has been known since its observation in H~{\sc i} by
\citet{1957AJ.....62...93K}. The presence of a North-South asymmetry in
stellar number density and velocity in the solar neighborhood has been found
in SDSS data \citep{2012ApJ...750L..41W}, and in velocity in the LAMOST survey
\citep{2013ApJ...777L...5C, 2015ApJ...801..105X}, and RAVE data
\citep{2013MNRAS.436..101W}. Beyond the solar neighborhood, low-latitude
overdensities have been observed, e.g., the Monoceros Ring
\citep{2002ApJ...569..245N, 2003MNRAS.340L..21I, 2003ApJ...594L.119C,
2014ApJ...791....9S, 2016ApJ...825..140M}, an overdensity in the direction of
the Triangulum and Andromeda galaxies \citep{2007ApJ...668L.123M,
2014ApJ...793...62S, 2015ApJ...801..105X, 2015MNRAS.452..676P} and A13
\citep{2010ApJ...722..750S, 2017ApJ...844...74L}, and in the direction of the
Galactic Center \citep{2014Natur.509..342F}. There is evidence that these
stellar populations originated from the disk \citep{2015MNRAS.452..676P,
2018ApJ...854...47S, 2018Natur.555..334B}. See, e.g.,
\citet{2018MNRAS.481..286L} for a discussion of these observations.

\begin{figure*}[ht!]
\begin{center}
\includegraphics[width=\textwidth]{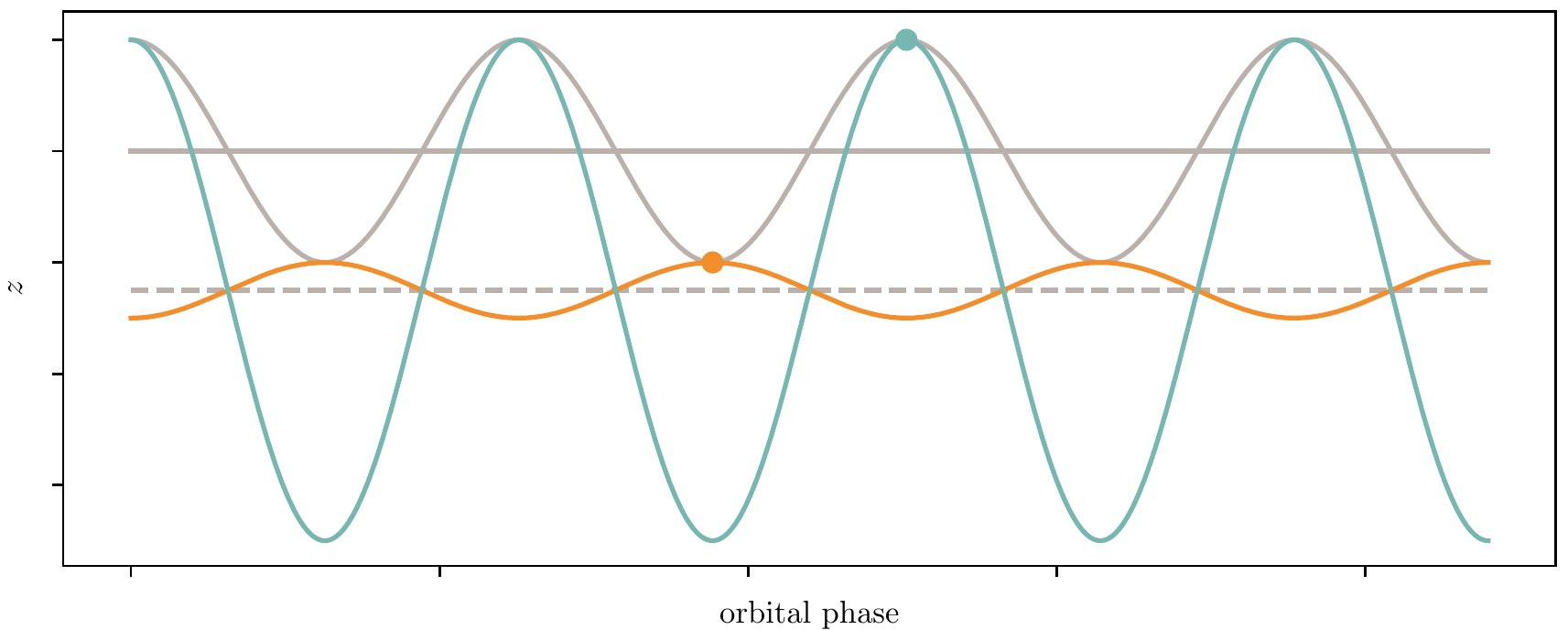}
\end{center}
\caption{Illustration showing the effect an error in the determination
of the coordinate midplane can have on orbit integration and action
estimation. The $x$-axis shows the orbital phase and the $y$-axis the vertical
height. The top gray curve depicts an example ``true'' orbit oscillating about
the true midplane (horizontal solid gray line). Consider an observer who
erroneously assumes the midplane is located at the horizontal dashed line.
Suppose this observer measures the phase-space position of the star at two
different orbital phases (teal and orange points). If the observer were to
then integrate the star's orbit using a model potential with the erroneous
midplane, they would obtain the teal and orange curves for the star's orbit,
respectively. The actions estimated from these two erroneous measurements
would subsequently differ, both from each other and from the true measurement
(in the potential with the correct coordinate system). Hence an incorrect
midplane in the potential model assumed will induce phase dependence in the
actions estimated for a given star in that potential.} \label{fig:cartoon}
\end{figure*}

With the vast improvement in the quality of phase-space measurements
due to \textit{Gaia} the assumption of axisymmetry is increasingly inadequate
\citep[e.g.,][]{2018Natur.561..360A, 2019MNRAS.485.3134L}. The presence of the
phase-space spiral itself is evidence of disk oscillations
\citep{2018Natur.561..360A}, but the fact that its strength changes across the
disk is further evidence \citep{2019MNRAS.485.3134L, 2019MNRAS.486.1167B}.
Even if the axisymmetric assumption were close enough for many purposes, the
parameters used in axisymmetric models of the Galactic potential may be biased
by the non-axisymmetries in the disk.

Such variations in the density of gas and stars cause the local midplane
position to vary as a function of radius and azimuth. Stars far from the Sun
have a local midplane that differs from our local midplane extrapolated onto
their position. Converting positions and velocities of more distant stars from
a heliocentric to a Galactocentric coordinate system thus introduces a
systematic bias in the $z$ coordinate. We show that this bias induces
non-Gaussian errors in the actions computed for these stars. The further from
the solar neighborhood the target star is, the more likely the mismatch will
result in large systematic uncertainty, especially in the vertical action
$J_z$. A similar argument applies to any remaining uncertainty in measurements
of the Galactocentric radius of the Sun, and to variations in the LSR.

In Section~\ref{sec:ref_frame}, we describe the general impact coordinate
system errors have on the measured actions. In Section~\ref{sec:local_fire},
we examine the azimuthal variations of the midplane itself in examples from
two classes of simulations: cosmological, hydrodynamical, zoom-in simulations
of isolated Milky Way-mass galaxies from the Feedback in Realistic
Environments (FIRE)
collaboration\footnote{\url{https://fire.northwestern.edu}}
\citep{2014MNRAS.445..581H, 2016ApJ...827L..23W, 2018MNRAS.480..800H}, and a
controlled N-body simulation of a Sagittarius encounter with a galaxy
otherwise tailored to the stellar mass, scale length, and scale height of the
Milky Way \citep{2018MNRAS.481..286L}. In Section~\ref{sec:discussion}, we
discuss the implications of midplane variations, and the resulting systematic
uncertainty in the vertical action, for action-space analyses. We also
estimate the expected midplane variations of the Milky Way based on the
observed velocity variations and three-dimensional dust maps. We summarize our
main results and conclude in Section~\ref{sec:conclusion}.

\section{Motivation} \label{sec:ref_frame}
We first demonstrate the significance to action computations of a systematic
offset in the determination of the Galactic midplane, distance to the Galactic
center, or LSR. We will find that such offsets are especially important for
disk-like orbits. The consequences we explore here may also arise from various
other systematic errors. For instance, the axisymmetric Galactic potential
model used in many works to compute actions may not be a good description of
the true potential\,---\,or the parameters used may yield a potential that is
systematically incorrect outside an original fitted region. In this work, we
assume that the Galaxy is perfectly described by our model axisymmetric
potential, and simply explore the consequences of offsets in the
Galactocentric coordinate system.

\subsection{Effect of Midplane Offset} \label{ssec:cartoon}
We present an illustration of an orbit in Figure~\ref{fig:cartoon} to show how
an inaccurate or erroneous determination of the midplane leads to a dependence
on orbital phase of the value of the actions calculated from a point in phase
space and an assumed potential model. The $x$-axis corresponds to orbital
phase and the $y$-axis to vertical height. The solid gray curve indicates the
true orbit of the star as it oscillates around the true midplane. The dashed
gray line, offset from the true midplane, is the midplane location used by an
observer to integrate the orbit of the star and estimate its actions. The
model potential is otherwise identical to the one in which the star is
actually moving.

Now suppose this observer makes a measurement of the star's position and
velocity at the teal point or the orange point (i.e., at two different orbital
phases). Then, the teal and orange curves correspond to the orbits that the
observer would compute for each point based on the potential model with the
offset midplane. In action space, this would correspond to a different value
of $J_z$ for the teal and orange points. In this way, assuming the wrong
coordinate system induces a phase-dependence in the actions estimated for the
star, which in the correct potential (in this example, the one with the
correct midplane) should be phase independent.

This example uses an offset in $z$, but analagous effects occur from offsets
in other coordinates, such as the distance to the Galactic center or the LSR.
A similar effect in which actions gain time-dependence due to a time-varying
potential was pointed out by \citet{2015A&A...584A.120B}.

In reality one only ever measures the position and velocity of a star
at a single instance in time. However, there are at least two possibilities
for going beyond a single temporal snapshot of the phase-space distribution.
If more than one star on the same or similar orbit could be identified (e.g.,
by identifying the remnants of a disrupted open cluster), then it would be
possible to constrain the disk potential in analogy to how stellar streams are
used to constrain the Galactic  potential \citep[e.g.][]{1999ApJ...512L.109J,
2001ApJ...551..294I, 2004ApJ...610L..97H, 2015ApJ...801...98S,
2015ApJ...803...80K, 2019MNRAS.483.1427L}. This would be challenging since, in
principle, one would want to use the stellar actions or other conserved
quantities, but as we discuss in this work they are subject to substantial
errors.  Another possibility would be to use chemical tagging to identify a
disrupted open cluster \citep{2002ARA&A..40..487F}, but this may be difficult
due to the large number of stars that are chemically indistinguishable though
presumably born in different clusters \citep{2018ApJ...853..198N}. A second
way to improve on a single temporal snapshot is to measure directly the
accelerations of stars \citep{2018arXiv181207581S}. While this would allow a
direct measurement of the aliasing effect in Figure~\ref{fig:cartoon}, the
observational challenges will likely limit this type of measurement to a small
number of local stars for the near future.

\subsection{Epicyclic Approximation} \label{ssec:epi_action}

Before turning to numerical methods, we derive analytic expressions for the
systematic error in the actions induced from offsets in the position (e.g.,
the midplane or Galactic center distance) or velocity (e.g., the LSR) of the
assumed coordinate system's origin.

We use the epicyclic approximation, which assumes that the motion in the $z$
and $R$ components of the orbit are decoupled and follow simple harmonic
motion about a circular and planar guiding orbit \citep[][Section~3.2 and
references therein]{2008gady.book.....B}. The radius of this orbit is referred
to as the guiding radius $R_g$. This approximation is an excellent description
of the thin disk and a good description of the thick disk in an axisymmetric
potential that ignores the influence of the Galactic bar and spiral arms. We
also make the assumption of a perfectly flat circular velocity curve with
$v_c(R) = v_c$, a good approximation near the solar circle
\citep[e.g.,][]{2017MNRAS.465...76M}.

Under this approximation, we can write down the cylindrical components
of the orbits as
\beq\label{eq:orbits_epi}
\begin{split}
R(t) &= R_g + A_R \sin{(\kappa t + \alpha)} \\
\phi(t) &= \Omega_c t \\
z(t) &= A_z \sin{(\nu t + \beta)}
\text{,}
\end{split}
\eeq
where $\kappa$ and $\nu$ are the radial/epicyclic and vertical frequencies, $\Omega_c
\equiv v_c/R_g$ is the orbital frequency of the guiding center, $A_R$ and
$A_z$ are the amplitudes in the radial and vertical coordinates, and $\alpha$
and $\beta$ are the initial orbital phases. Similarly, the velocities of the
orbit are given by:
\beq\label{eq:orbits_vel_epi}
\begin{split}
v_R(t) &= \kappa A_R \cos{(\kappa t + \alpha)} \\
v_{\phi}(t) &= v_c \\
v_z(t) &= \nu A_z \cos{(\nu t + \beta)}
\text{.}
\end{split}
\eeq

In this case, 
the azimuthal action is
\citep[][Section~3.5.3b]{2008gady.book.....B}:
\beq\label{eq:Jphi_epi}
J_{\phi} = R_g v_c\text{,}
\eeq
and the radial and vertical actions are
\beq\label{eq:JR_Jz_epi}
\begin{split}
J_R &= E_R / \kappa \\
J_z &= E_z / \nu\text{,} 
\end{split}
\eeq
where $E_R$ and $E_z$ are the energy per unit mass in the radial and vertical
coordinates, respectively. Therefore,
\beq\label{eq:JR_Jz_epi_energy}
\begin{split}
J_R &= \frac{v_R^2 + \kappa^2 (R-R_g)^2}{2\kappa} \\
J_z &= \frac{v_z^2 + \nu^2 z^2}{2\nu}\text{.}
\end{split}
\eeq
Using Equations~\eqref{eq:orbits_epi} and \eqref{eq:orbits_vel_epi}, we can
simplify this:
\beq\label{eq:JR_Jz_epi_final}
\begin{split}
J_R &= \frac{\kappa A_R^2}{2} = \frac{v_{R,\text{max}}^2}{2\kappa} \\
J_z &= \frac{\nu A_z^2}{2} = \frac{v_{z,\text{max}}^2}{2\nu}\text{,}
\end{split}
\eeq
where the last equality in each line comes from the fact that
$v_{R,\text{max}} = \kappa A_R$ and $v_{z,\text{max}} = \nu A_z$.

Notice that while the value for each of $J_{\phi}$, $J_R$, and $J_z$ is phase
independent, the contribution from the kinetic and potential terms in
Equation~\eqref{eq:JR_Jz_epi_energy} is phase dependent. Now assume that the
coordinates $(R, z, v_{\phi}, v_R, v_z)$ are offset by $(\Delta R, \Delta z,
\Delta v_{\phi}, \Delta v_R, \Delta v_z)$. We can then apply the standard
propagation of errors formula to Equation~\eqref{eq:JR_Jz_epi_energy} to
determine the error in each of the actions. 
For $J_R$, the induced error is:
\beq\label{eq:induced_JR}
\frac{\Delta J_R}{J_R} = \frac{2(R-R_g)}{A_R^2}\Delta R
                         + \frac{2v_R}{v_{R,\text{max}}^2} \Delta v_R \text{.}
\eeq
For $J_{\phi}$, the induced error
is:
\beq\label{eq:induced_Jphi}
\frac{\Delta J_{\phi}}{J_{\phi}} = \frac{\Delta R}{R_g}
                                    + \frac{\Delta v_{\phi}}{v_c}\text{.}
\eeq
For $J_z$, the induced error is:
\beq\label{eq:induced_Jz}
\frac{\Delta J_z}{J_z} = \frac{2z}{A_z^2}\Delta z
                         + \frac{2v_z}{v_{z,\text{max}}^2} \Delta v_z \text{.}
\eeq
We have ignored second order contributions.

\begin{deluxetable*}{cCCCCCCCCC}
\tablecaption{Names and properties of the three orbits considered in this work, where
$z_{\text{max}}$ is the maximum height of the orbit,
$\frac{1}{2}(R_{\text{max}} - R_{\text{min}})$ is the magnitude of the radial
excursions of the orbit, and $\kappa$ and $\nu$ are the radial/epicyclic and
vertical frequencies of the orbit. In the epicyclic approximation, $A_z =
z_{\text{max}}$ and $A_R = \frac{1}{2}(R_{\text{max}} - R_{\text{min}})$.
\label{tab:orbits}} \tablehead{\colhead{name} & \colhead{\makecell{initial \\ position}} &
\colhead{\makecell{initial \\ velocity}} & \colhead{$J_R$} & \colhead{$J_{\phi}$} &
\colhead{$J_z$} & \colhead{$z_{\text{max}}$} &
\colhead{$\frac{1}{2}(R_{\text{max}} - R_{\text{min}})$} & \colhead{$\kappa$} &
\colhead{$\nu$} \\ \colhead{ } & \colhead{$(\mathrm{kpc})$} &
\colhead{$(\mathrm{km\,s}^{-1})$} & \colhead{$(\mathrm{kpc\,km\,s^{-1}})$} &
\colhead{$(\mathrm{kpc\,km\,s^{-1}})$} & \colhead{$(\mathrm{kpc\,km\,s^{-1}})$} &
\colhead{$(\mathrm{kpc})$} & \colhead{$(\mathrm{kpc})$} &
\colhead{$(\mathrm{Myr}^{-1})$} & \colhead{$(\mathrm{Myr}^{-1})$}}
\startdata
\thin{} & (8, 0, 0) & (0, -190, 10) & 40 & -1500 & 0.69 & 0.12 & 1.3 & 0.049
& 0.093 \\
\thick{} & (8, 0, 0) & (0, -190, 50) & 33 & -1500 & 23 & 0.85 &
1.2 & 0.048 & 0.061 \\
\halo{} & (8, 0, 0) & (0, -190, 190) & 33 & -1500 &
530 & 6.2 & 2.3 & 0.033 & 0.025
\enddata
\end{deluxetable*}

Since most of the time stars will be at maximum amplitude (i.e., turnaround)
in both $R$ and $z$, we can approximate the order of magnitude of the
systematic error in the actions by
\beq\label{eq:Ji_err_mosttime}
\begin{split}
\frac{\Delta J_{R}}{J_{R}} &= \frac{2\Delta R}{A_R} \\
\frac{\Delta J_{\phi}}{J_{\phi}} &= \frac{\Delta R}{R_g}
                                    + \frac{\Delta v_{\phi}}{v_c} \\
\frac{\Delta J_{z}}{J_{z}} &= \frac{2\Delta z}{A_z} \text{,}
\end{split}
\eeq
where we have again ignored second order terms.

Because the energies in the radial and vertical coordinates are
related to the radial and vertical actions by a constant
(Equation~\eqref{eq:JR_Jz_epi}), the fractional error in these two energies is
the same as the actions.\footnote{This ignores any potential error in the
epicyclic and vertical frequencies.} The energy in the azimuthal coordinate
can be readily computed ($E_{\phi} = \frac{1}{2}v_{\phi}^2$). Using standard
propagation of errors, we see that $\Delta E_{\phi}/E_{\phi}$ and $\Delta
J_{\phi}/J_{\phi}$ (Equation~\eqref{eq:induced_Jphi}) differ by a $\Delta
R/R_g$ term. Since $R_g\approx 8\,\kpc$, we expect this term to be negligible
and we therefore conclude that:
\beq\label{eq:error_total_energy}
\frac{\Delta E}{E} \approx \frac{\Delta J_R}{J_R} + 2\frac{\Delta J_{\phi}}{J_{\phi}}
                     + \frac{\Delta J_z}{J_z}\text{.}
\eeq

In the remainder of this section, we compare our analytic estimates of the
effect of a midplane offset on actions against numerical calculations. A
numerical evaluation of the effect of velocity offsets on actions is deferred
to future work, as we discuss in Section~\ref{ssec:lsr_var}.

\subsection{Numerical Methods} \label{ssec:action_comp}
We now quantify the argument made in Section~\ref{ssec:cartoon} using
numerical computations of the actions for a range of orbits in a model
Galactic potential. We compute actions as in \citet{2018ApJ...867...31B},
using the code \texttt{gala}~v0.3 to perform orbit integrations and conversion
to action space \citep{gala, gala:zenodo}. To compute actions we use the
torus-mapping technique first presented by \citet{1990MNRAS.244..634M} and
adapted by \citet{2014MNRAS.441.3284S} to calculate actions for an orbital
time-series starting from a phase-space position $(x, v)$ and integrated in a
potential $\Phi$. For our Galactic potential we use \texttt{MWPotential},
based on the Milky Way potential available in \texttt{galpy}
\citep{2015ApJS..216...29B}, which includes a Hernquist bulge and nucleus
\citep{1990ApJ...356..359H}, a Miyamoto--Nagai disk
\citep{1975PASJ...27..533M}, and a Navarro, Frenk, \& White
\citeyear{1997ApJ...490..493N} halo, and is fit to empirically match some
observations. We use the Dormand-Prince 8(5,3) integration scheme
\citep{Dormand80:integrator} with a timestep of $1\,\Myr$ and integrate for
$5\,\Gyr$, corresponding to $\sim 20$ orbits for a Sun-like star.

We assume the Sun is located at $(8.2, 0, 0)\,\kpc$. None of our orbit
integrations depend on the value of the LSR in this toy potential (though this
is important when using real data, since the conversion from heliocentric to
Galactocentric coordinates depends on the LSR). In this potential, we have
that the circular velocity $v_{\text{circ}}$ is $231\,\kms$ at the solar
circle.

Other methods for computing actions are used in the literature. For example,
the St\"ackel Fudge method \citep{2016MNRAS.457.2107S}, which uses a single
St\"ackel potential (with analytic actions) to approximate the Galactic
potential \citep{1985MNRAS.216..273D, 2012MNRAS.426.1324B}, was used in many
recent works exploring actions in the Galactic disk
\citep[e.g.,][]{2019MNRAS.484.3291T, 2018MNRAS.481.4093S,
2018arXiv180803278T}. For disk-like orbits, existing implementations of the
St\"ackel Fudge method are of acceptable accuracy, but since we also consider
halo-like orbits in this work (where the St\"ackel Fudge method is inaccurate)
we choose to use orbit integration and torus mapping throughout
\citep{2016MNRAS.457.2107S}.

\subsection{Quantification of the Midplane Effect} \label{ssec:quant}

\begin{figure*}[ht!]
\begin{center}
\includegraphics[width=\textwidth]{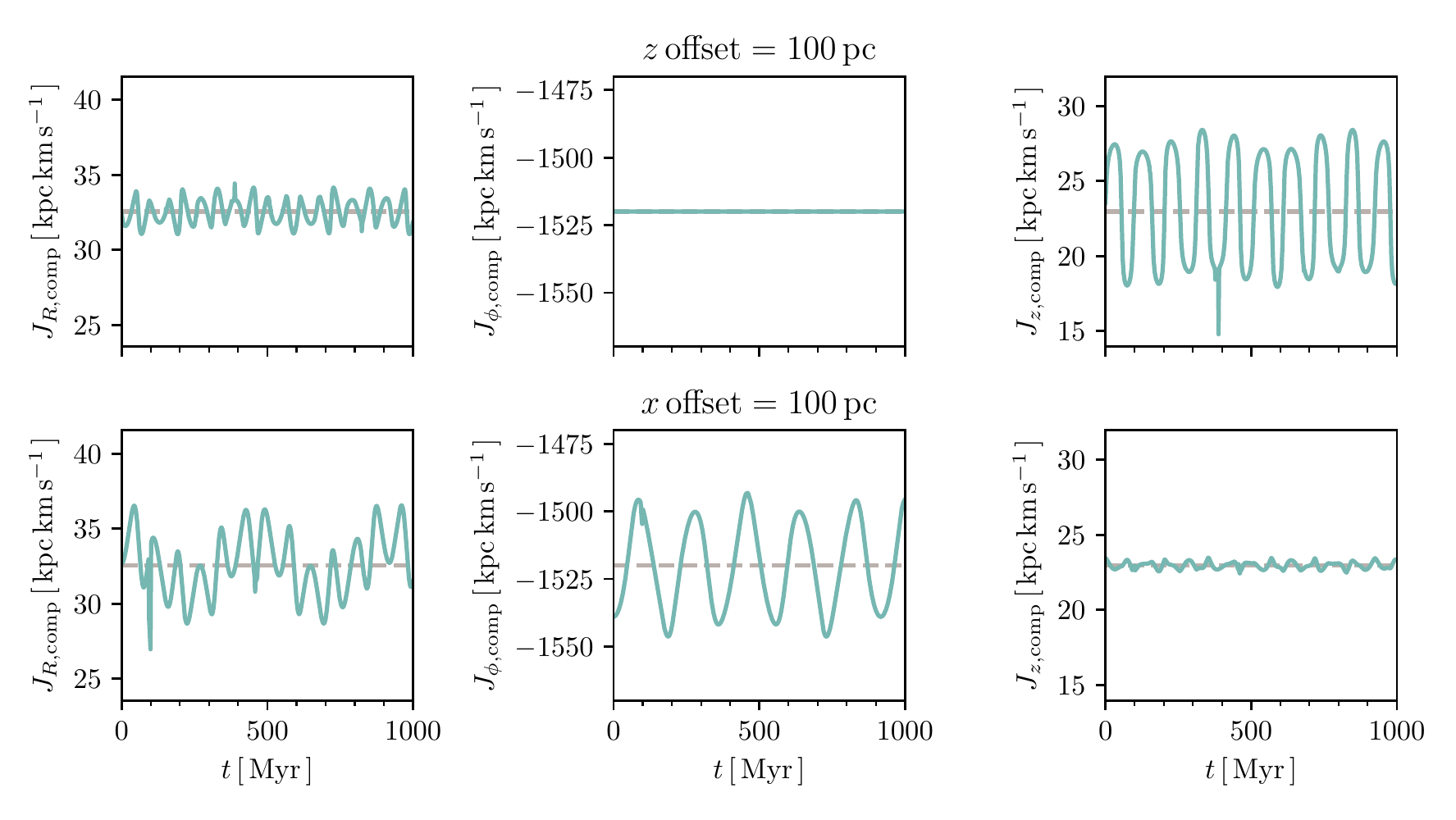}
\end{center}
\caption{The artificial phase-dependence in the computed actions with an
error in the Galactocentric coordinate system. We consider here \thick{},
which has actions of $(J_R, J_{\phi}, J_z) = (38, -1500, 7.0)\,\actunit$ and
$z_{\text{max}}=850\,\pc$ (see Table~\ref{tab:orbits}). We integrate the orbit
according to the procedure laid out in Section~\ref{ssec:action_comp}, and
which we plot in Appendix~\ref{app:orbits}. Then, we subtract $100\,\pc$ from
the $z$ value (upper panels) or the $x$ value (lower panels) of each position
in the orbit, corresponding to an erroneous observer assuming a midplane
(upper) or solar radius (lower) that is off by $100\,\pc$. We then allow an
observer to measure the orbit over $1\,\Gyr$ and perform the same orbit
integration procedure at each timestep, and report the values of the actions,
with the true values given as horizontal dashed lines. The computation of
$J_{\phi}$ is pristine to errors in $z$, with only numerical artifacts
remaining. Only small errors are induced in $J_R$, with the middle $90\%$ of
values over the $\Gyr$ being within $\sim8\%$ of the true $J_R$. As expected,
large errors are induced in $J_z$ with a $100\,\pc$ offset in $z$, with the
middle $90\%$ of values being within $\sim43\%$ of the true $J_z$. The $x$
offset induces uncertainties in $J_R$, $J_{\phi}$, and $J_z$ of $\sim21\%$,
$\sim3\%$, and $\sim3\%$.}
\label{fig:one_orbit_wrong_ref}
\end{figure*}

\begin{figure}
\begin{center}
\includegraphics[width=\columnwidth]{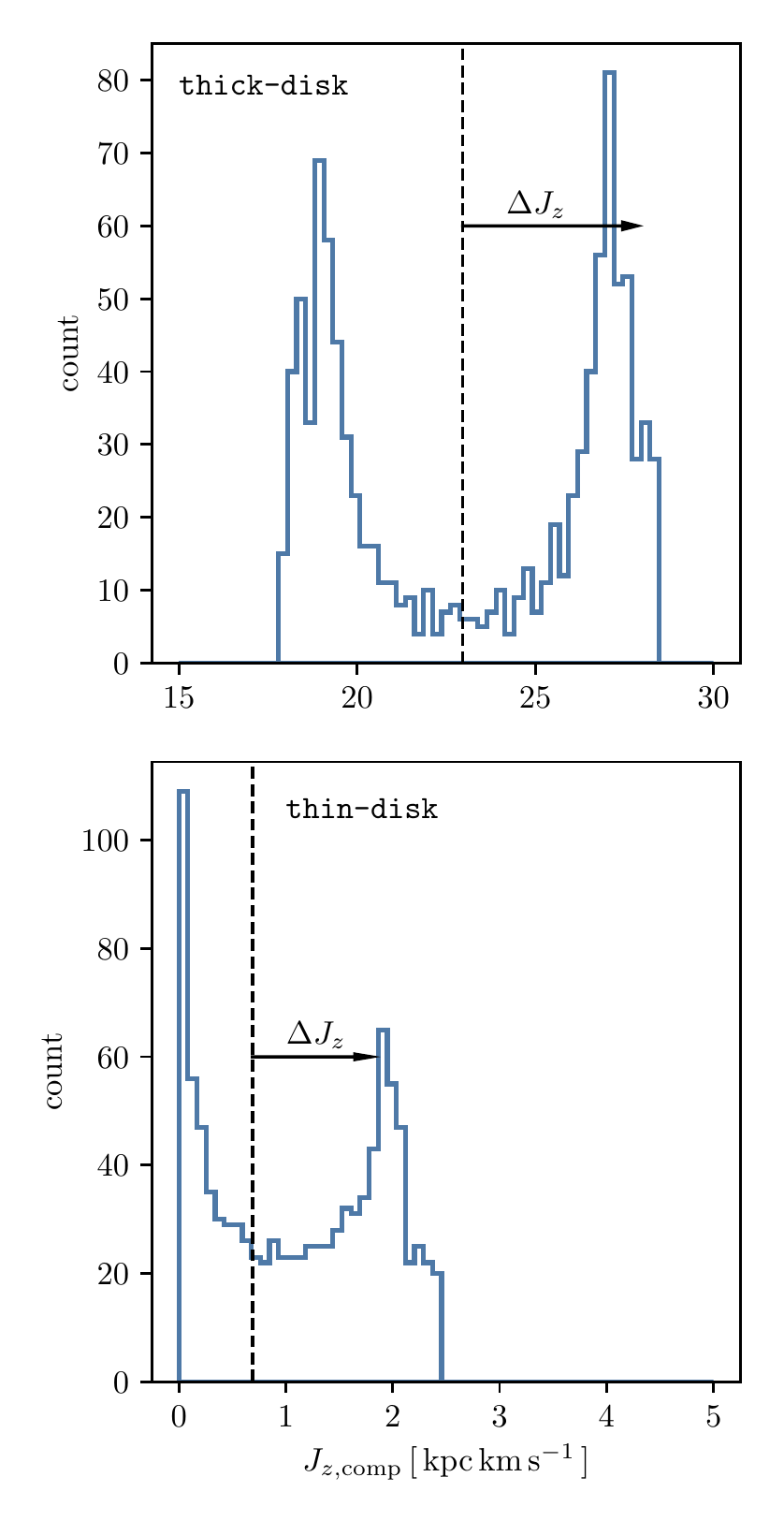}
\end{center}
\caption{A histogram of the computed values of $J_z$ at different orbital
phases for \thick{} (top panel) and \thin{} (bottom panel) assuming a $z$
offset of $100\,\pc$. One can see that if the observed $z$ values have a bias
(from, e.g., an incorrectly computed midplane), then the induced error
distribution in $J_z$ is decidedly non-Gaussian. Therefore, any sort of error
propagation must take this into account. A heuristic explanation for the shape
of each panel is given in the text. We also plot one half the 95th percentile
minus the 5th percentile of each distribution as a horizontal arrow anchored
on the true $J_z$ value. We call this $\Delta J_z$ and will use it (along with
the similarly defined $\Delta J_R$ and $\Delta J_{\phi}$) to empirically
describe the error distribution. We see that $\Delta J_z$ roughly corresponds
to the distance from the true $J_z$ value to one of the modes of the
distribution of computed $J_z$ values. Similar plots for $J_R$ induced by a
$z$ offset and $J_R$ and $J_{\phi}$ induced by an $x$ offset are given in
Appendix~\ref{app:hist}.}
\label{fig:Jz_hist}
\end{figure}

We now quantify how a systematic error in the Galactocentric coordinate system
induces phase-dependence in the actions calculated from the observed position
and velocity of a star. We consider three orbits in the model potential
described in Section~\ref{ssec:action_comp} that are typical of stars in the
thin disk, thick disk, and halo. We summarize their initial positions in phase
space, the actions computed by integrating their orbits in the correct
potential, and other properties in Table~\ref{tab:orbits}. Each orbit,
integrated without systematic coordinate errors, is plotted in
Appendix~\ref{app:orbits}. We will refer to these orbits by their names
(\thin{}, \thick{}, \halo{}) henceforth.

We begin with \thick{}. Consider an observer who can measure the orbit's
phase-space position at many different times (and hence different orbital
phases), but does so using a coordinate system in which the midplane is
systematically offset in height by $100\,\pc$ from its actual location. To
model this we subtract the vector $(0, 0, 100)\,\pc$ from each position in the
orbit. This corresponds to an observer physically located at, e.g., the
position $(8, 0, 0)\,\kpc$ in the coordinate system of the true potential, but
erroneously thinking they are located at $(8, 0, 0.1)\,\kpc$.

We consider the observer making a measurement, integrating an orbit, and
computing actions every megayear using the prescription above. However, we
specify the star's starting position using the systematically offset
coordinate system. Essentially we are shifting and then reintegrating at each
point along the original orbit. The actions computed using the offset
coordinate system for each phase-space starting point are shown for the first
gigayear of the orbit in the upper panels of
Figure~\ref{fig:one_orbit_wrong_ref}.\footnote{Occasionally the numerical
scheme fails and very large actions are reported by \texttt{gala}\,---\,we
perform a $4\sigma$ clip on each action to exclude such orbits, but this only
excludes a total of $5$ orbits out of the $1000$ considered for
Figure~\ref{fig:one_orbit_wrong_ref}. Some numerical artifacts remain, but the
vast majority of orbits are computed properly.}

We also perform the same procedure in the lower panels but assuming an $x$
component offset of $100\,\pc$, i.e., subtracting the vector $(100, 0,
0)\,\pc$. This is equivalent to a measurement error in the distance from the
Sun to the Galactic center.

Figure~\ref{fig:one_orbit_wrong_ref} shows that the actions computed in the
offset coordinate systems oscillate as a function of the time/orbital phase at
which the star's phase-space position is observed. This time dependence comes
even though the observer is using the correctly constructed, best-fit, static,
axisymmetric potential. The relative size of the phase variation in each
action depends on the direction of the systematic offset as well as the true
values of the actions (i.e. the type of orbit). In reality, we will have one
measurement of the phase-space position to work with, in which case the
determination of the orbital phase in $R$ or $z$ is degenerate with the degree
of systematic offset in that coordinate (see Figure~\ref{fig:cartoon}). In the
following we therefore quote percentile ranges for the possible values
computed for each action as a proxy for the effect of these systematic errors
in the coordinate system.

For a systematic offset in $z$ (upper panels), the 95\textsuperscript{th}
minus 5\textsuperscript{th} percentiles are $2.2$ and $6.2\,\actunit$ for
$J_R$ and $J_z$, respectively. As a fraction of the true values, these are
$5.7\%$ and $86\%$, respectively. The spread induced in $J_{\phi}$ is
negligible, as expected since $J_{\phi}$ only depends on the $x$- and
$y$-components of the position and velocity of the stars.\footnote{In
practice, however, $J_{\phi}$ is computed as part of the torus-fitting
method.} It is worth pointing out that a $100\,\pc$ offset in an orbit with
$z_{\text{max}}=850\,\pc$\,---\,a $12\%$ error\,---\,induced an $86\%$ spread
in the computation of $J_z$.

For a systematic offset in $x$ (or distance to the Galactic center), the
95\textsuperscript{th} minus 5\textsuperscript{th} percentiles are $6.9$,
$47$, and $0.71\,\actunit$ for $J_R$, $J_{\phi}$ and $J_z$, respectively.
These are fractionally $21\%$, $3.1\%$ and $3.1\%$ in these actions,
respectively, despite only a $1.2\%$ error in the distance to the Galactic
center.

In Figure~\ref{fig:Jz_hist}, we plot a histogram of the values of $J_z$
computed at different orbital phases for \thick{} (top panel) and \thin{}
(bottom panel), assuming a $z$ offset of $100\,\pc$ (as in the upper right
panel of Figure~\ref{fig:one_orbit_wrong_ref}). The true value is plotted as a
vertical dashed line. The systematic error in $J_z$ induced by a systematic
offset in $z$ is non-Gaussian and bimodal; neither of the modes is centered on
the null value. In the case of \thin{} (bottom panel), we see that, in
addition to the prior complications, the distribution is not even centered on
the true value. This comes about when the midplane error is~$\gtrsim
z_{\text{max}}$, where $z_{\text{max}}$ is the maximum height of the orbit
(equivalent to $A_z$ in the epicyclic approximation, see
Section~\ref{ssec:epi_action}).

\begin{figure*}[htb!]
\begin{center}
\includegraphics[width=\textwidth]{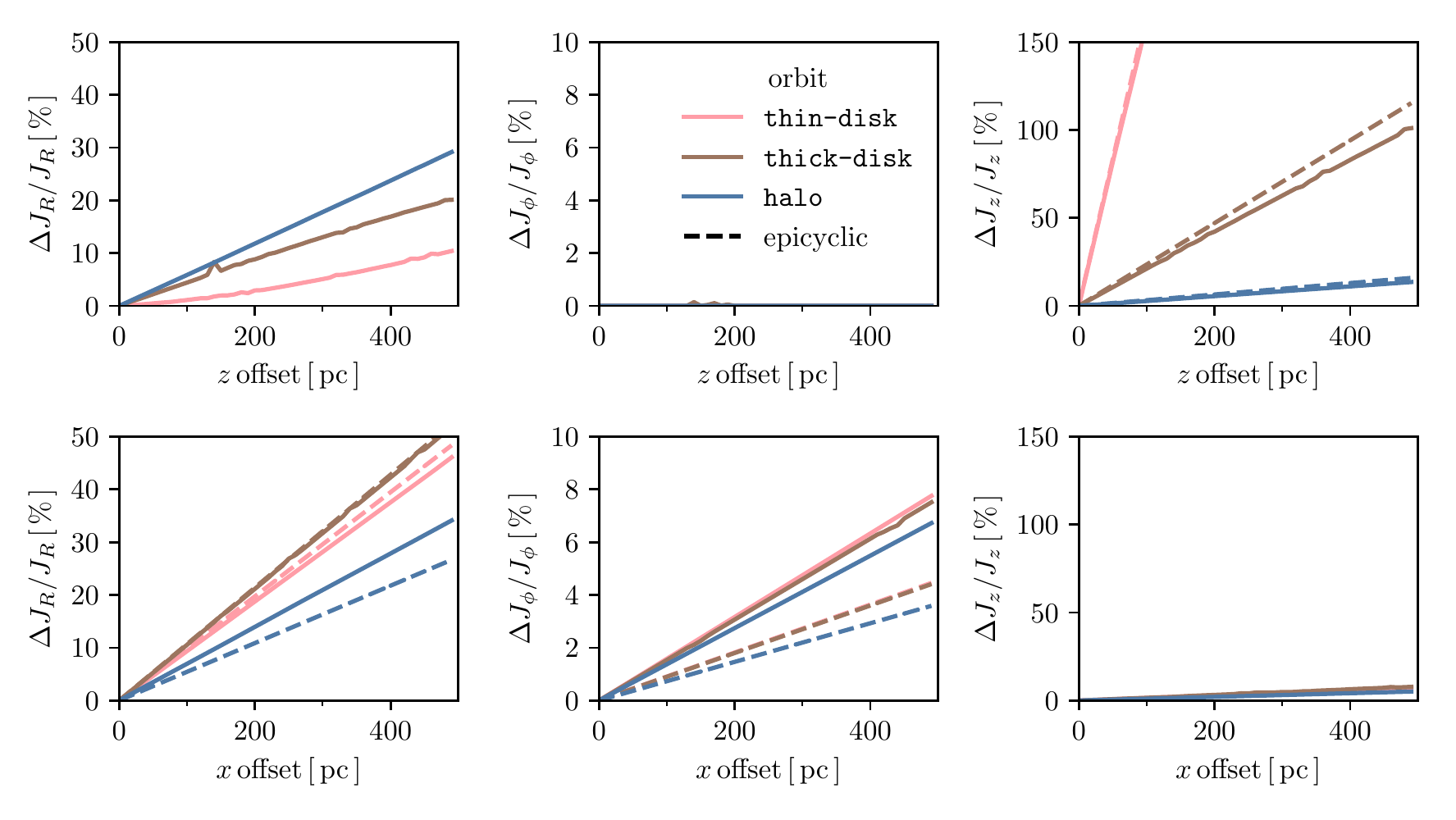}
\end{center}
\caption{We report the fractional error in each action $J_i$ induced by
coordinate system offsets for \thin{}, \thick{}, and \halo{}
(Table~\ref{tab:orbits}). The error ($\Delta J_i$) is computed as one half the
95th minus 5th percentile of the distribution of computed action values. See
discussion in the text and Figure~\ref{fig:Jz_hist} for the justification in
using this to measure the magnitude of the induced error. The left, center,
and right panels show the result for $J_R$, $J_{\phi}$, and $J_z$,
respectively. The upper panels consider an offset in $z$ and the lower panels
consider an offset in $x$ (equivalently, an offset in the solar radius). In
some panels, we also plot as dashed lines the epicyclic prediction of the
induced action error (Equation~\eqref{eq:Ji_err_mosttime}). In the epicyclic
approximation, a $z$ offset only induces an error in $J_z$\,---\,for all three
orbits the epicyclic approximation is a good description of the $J_z$ error.
An $x$ offset induces an error in $J_R$ and $J_{\phi}$. The error in $J_R$ is
somewhat well-described for \thin{} and \thick{}, and a poor description for
\halo{}. For $J_{\phi}$, the epicyclic approximation is not a good description
for any orbit.}
\label{fig:many_orbit_wrong_ref}
\end{figure*}

In Appendix~\ref{app:hist} we plot the same histogram as in
Figure~\ref{fig:Jz_hist}, but for the distributions of $J_R$ induced by a $z$
offset (upper left panel of Figure~\ref{fig:many_orbit_wrong_ref}) and the
distributions of $J_R$ and $J_{\phi}$ errors induced by an $x$ offset (lower
left and lower center panels of Figure~\ref{fig:many_orbit_wrong_ref},
respectively). We find similar error distributions as in
Figure~\ref{fig:Jz_hist}, with the exception that the computed $J_R$
distribution induced by an $x$ offset more closely resembles a Gaussian
distribution.

We now suggest a heuristic explanation for the shape of
Figure~\ref{fig:Jz_hist}. Consider first the \thick{} (top panel), where the
offset in $z$ is much less than the $z_{\text{max}}$ of the orbit. The peaks
in the distribution correspond to the turning points of the orbit (or points
of maximum vertical amplitude), where $v_z \sim 0$ and where the star is on
most of it orbit. This is why the distribution, which is calculated at evenly
spaced time intervals, peaks at these values. For \thin{} (bottom panel), the
offset in $z$ is comparable to $z_{\text{max}}$. Now, there will be some
points in the orbit where $v_z = 0$ and $z=0$ (in the erroneous coordinate
system). At these points, the computed $J_z$ will vanish. The asymmetry and
systematic offset then comes about because of the constraint that $J_z \geq
0$.\footnote{This argument is similar to ones given in cosmology for why
gravity produces non-Gaussianity in the density field, since the density
cannot become negative but it can grow arbitrarily large.}

Gaussian summary statistics are clearly insufficient to describe the
distribution shown in Figure~\ref{fig:Jz_hist}. We therefore elect to measure
this error by computing one half the 95th percentile minus the 5th percentile
of the distribution of action values. We refer to this quantity as $\Delta
J_i$ for each action and plot it in Figure~\ref{fig:Jz_hist} as a horizontal
arrow anchored on the true action value. Because of the bimodality of the
error distribution, this quantity roughly measures the distance from the true
action value to the peak of one of the modes. Furthermore, this bimodality
also implies that $\Delta J_i$ is not very sensitive to the exact percentiles
used. This summary statistic does not reflect the bias induced when the
midplane error is~$\gtrsim z_{\text{max}}$.

We now repeat the same procedure as in Figure~\ref{fig:one_orbit_wrong_ref}
but for systematic offsets between $0$ and $500\,\pc$ in the $z$ and $x$
components. In Figure~\ref{fig:many_orbit_wrong_ref}, we report $\Delta
J_i/J_i$ for the three different fiducial orbits in Table~\ref{tab:orbits}.
The upper panels of Figure~\ref{fig:many_orbit_wrong_ref} shows the spread
induced in each action for an offset in the $z$-component. In the lower panels
we consider offsets in the $x$ component (i.e. the solar radius). The
left, center, and right columns show the fractional spread in the values
computed for $J_R$, $J_{\phi}$, and $J_z$, respectively.

In the upper middle panel of Figure~\ref{fig:many_orbit_wrong_ref}, there is
essentially no spread in the determination of $J_{\phi}$. This is expected
since $J_{\phi}$ is independent of $z$ and is thus unaffected by offsets in
$z$, as discussed earlier. Indeed, the result we found in
Figure~\ref{fig:one_orbit_wrong_ref} for \thick{} holds for all orbit types.
This is also a demonstration of the robustness of the integration and action
calculation methods we use.

The upper right panel of Figure~\ref{fig:many_orbit_wrong_ref} shows that the
fractional error in $J_z$ is more exaggerated for more planar (disk-like)
orbits. For \thin{}, a systematic offset of $15\,\pc$ in the $z$-coordinate
results in a $25\%$ deviation in $J_z$, while a $120\,\pc$ offset results in
the same deviation for \thick{}. We find that \halo{} is relatively
insensitive to errors in the midplane, with only $\sim15\%$ error in
$J_z$ out to an offset of $500\,\pc$.

For the offset in the solar radius (lower panels), the error is largest for
$J_R$, with some deviations resulting in $J_{\phi}$ and relatively small
deviations in $J_z$. In the lower center and lower right panels all three
lines nearly overlap.

In each panel of Figure~\ref{fig:many_orbit_wrong_ref}, where relevant, we
include the estimation of the action errors derived under the epicyclic
approximation from Equation~\eqref{eq:Ji_err_mosttime}, with $\Delta
v_{\phi}=0$, as dashed lines in the color of each orbit. This equation is
relevant since during most of the orbit the star will be close to maximum
radial and vertical amplitude. Note that we consider an error in the
$x$-coordinate $\Delta x$, which is not exactly the same as $\Delta R$. For
observations of stars close to us, we have that $\Delta x \sim \Delta R$, but
for the experiment performed in this section we consider observations of the
star throughout its entire orbit. This introduces a factor of $2/\pi$ when
converting from $\Delta x$ to $\Delta R$, which we derive in
Appendix~\ref{app:deltax}.

The epicyclic approximation is a good predictor of $\Delta J_z$, even for \halo{}.
It performs similarly for $\Delta J_R$, now underpredicting for \halo{}
and slightly overpredicting for \thin{}. Note that for the
particular orbits we chose, \thin{} has slightly larger $A_R$ than
\thick{}, and so we actually expect the epicyclic approximation to perform
slightly worse for \thin{} in this case. The epicyclic approximation
underpredicts $\Delta J_{\phi}$ for all orbits.

\begin{figure}
\begin{center}
\includegraphics[width=\columnwidth]{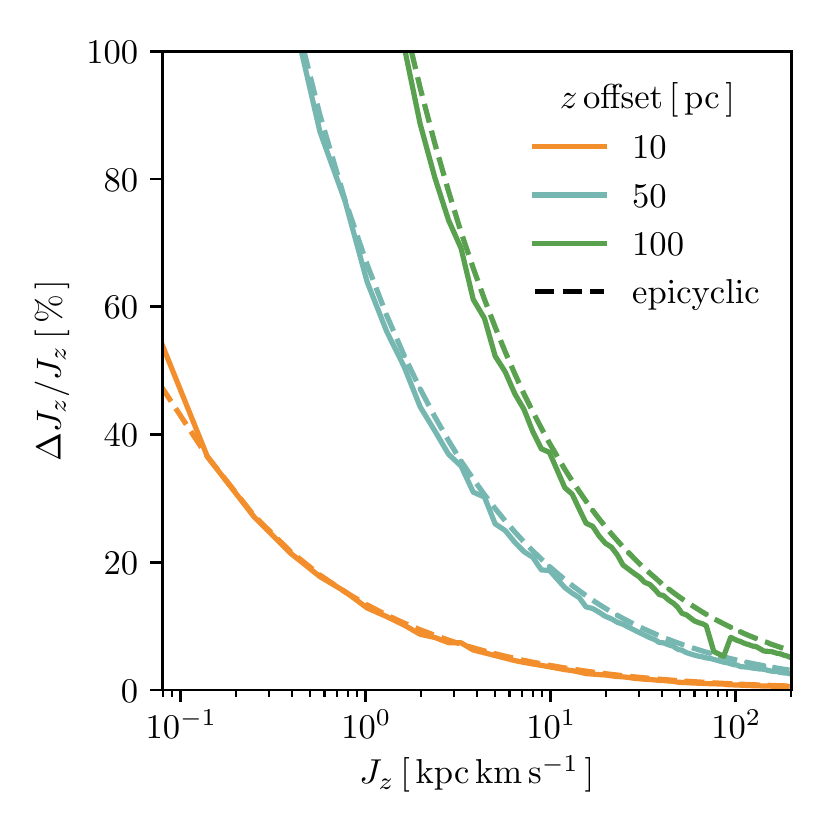}
\end{center}
\caption{The fractional error in $J_z$ as a function of $J_z$ for a few
different offsets in $z$. All orbits have the same initial position of $(8, 0,
0)\,\kpc$ and velocity $(0, -190, v_z)\,\kms$, where we vary
$v_z$.\footref{note:vz_orbits} We show this for a $z$ offset of $10$, $50$,
and $100\,\pc$ (orange, teal, and green, respectively). As before, the error
($\Delta J_z$) is one half the 95th minus 5th percentile of the distribution
of computed $J_z$ values over the course of the orbit. There are large errors
for \thin{}-like orbits ($J_z \sim 0.7\,\actunit$), even for a small midplane
offset of $10\,\pc$. As dashed lines in each color we also plot the prediction
for $\Delta J_z/J_z$ from the epicyclic approximation
(Equation~\eqref{eq:Ji_err_mosttime}), which shows excellent agreement with
the numerically computed values.}
\label{fig:dJz_fun_Jz}
\end{figure}

To further understand the effect of the midplane error, we also plot the
fractional error in $J_z$ as a function of $J_z$ for $z$ offsets of $10$,
$50$, and $ 100\,\pc$ (orange, teal, and green, respectively) in
Figure~\ref{fig:dJz_fun_Jz}. For each orbit, we set the initial position to be
$(8,0,0)\,\kpc$ and the initial velocity to be $(0, -190, v_z)\,\kms$, where
we vary $v_z$.\footnote{\label{note:vz_orbits}One can recover \thin{},
\thick{}, and \halo{} by setting $v_z=10$, $50$, and $190$, respectively,
giving $J_z \simeq 0.7, 20,$ and $500\,\actunit$ (see
Table~\ref{tab:orbits}).} For a \thin{}-like orbit ($J_z\sim0.7\,\actunit$),
even a $10\,\pc$ offset in $z$ is enough to induce a $\sim20\%$ error in
$J_z$. For larger values of $J_z$, the fractional errors are suppressed, but
the induced error can still be large depending on how great the $z$ offset is.
We also plot the epicylic prediction for $\Delta J_z / J_z$ from
Equation~\eqref{eq:Ji_err_mosttime} as dashed lines for each $z$ offset. We
find that the epicyclic approximation matches the numerical estimate quite
well.

\section{Azimuthal Midplane Variations}
\label{sec:local_fire}

The stellar midplane of the Galaxy should vary as a function of azimuth and
Galactocentric radius due to small, local variations in the stellar density.
Hints of this variation as a function of Galactocentric radius have been noted
through their impact on the stellar velocity distribution pre-\textit{Gaia} by
\citet{2012ApJ...750L..41W}, \citet{2013ApJ...777L...5C}, and
\citet{2013MNRAS.436..101W} and recently post-\textit{Gaia} by
\citet{2019arXiv190209569F}. As pointed out by, e.g.,
\citet{2014ApJ...797...53G} and \citet{2019ApJ...871..145A}, among many
others, the gas distribution in the Galaxy also shows significant density
variation across the disk.

The local Galactocentric coordinate system is defined based on the location of
the Sun relative to the midplane. Extending this coordinate system to a global
one therefore introduces systematic errors in the $z$ components of stellar
positions. As discussed in Section~\ref{sec:ref_frame}, this systematic error
introduces errors in integrating orbits and computing actions.

We specifically consider \emph{azimuthal} variations in the midplane at the
solar circle, as defined by the stellar mass density. Since, to our knowledge,
there are no direct empirical measurements of these variations in the Milky
Way, we use example simulations from two classes of simulations to estimate
the size of this effect.

One set are three zoom-in, cosmological hydrodynamical simulations of isolated
Milky Way-mass galaxies from the FIRE collaboration, described briefly in
Section~\ref{ssec:cosmozoom}. These include stars, gas, and dark matter in a
fully cosmological setting but are not tailored to specific properties of the
Milky Way (such as the scale height or scale length, or the details of the
accretion history). We use these simulations to span the range of
possibilities for azimuthal midplane variations.

\begin{deluxetable*}{cCCCCC}[htb!]
\tablecaption{Stellar and gas disk scale heights of the Milky Way and the
FIRE galaxies considered in this work (described in
Section~\ref{ssec:cosmozoom}). For comparison, we also give the median
softening lengths for the FIRE galaxies, computed for cold gas
($T<1000\,\text{K}$) and stars with $\abs{R-R_0} < 0.5\,\kpc$ and $\abs{z} <
1\,\kpc$. We have assumed that $R_0=8.2\,\kpc$.\label{tab:scale_height}}
\tablehead{\colhead{galaxy} & \colhead{\makecell{cold\tablenotemark{a} gas disk \\ scale height}} &
\colhead{\makecell{thin disk \\ scale height}} & \colhead{\makecell{thick disk \\ scale height}} & \colhead{\makecell{cold\tablenotemark{a} gas \\
softening length}} & \colhead{\makecell{stellar \\ softening length}} \\
\colhead{} & \colhead{(pc)} & \colhead{(pc)} & \colhead{(pc)} & \colhead{(pc)}
& \colhead{(pc)} }
\startdata
Milky Way\tablenotemark{b} & 40 & 300 & 900 & \nodata & \nodata \\
\mi\tablenotemark{c} & 800\tablenotemark{d} & 480 & 2000 & 53.4 & 11.2 \\
\mf\tablenotemark{c} & 360 & 440 & 1280 & 57.2 & 11.2 \\
\mm\tablenotemark{c} & 250 & 290 & 1030 & 60.1 & 11.2 \\
\enddata
\tablenotetext{a}{$T<100\,\text{K}$}
\tablenotetext{b}{\citet{2008ApJ...673..864J,2016ARAA..54..529B}}
\tablenotetext{c}{\citet{2018arXiv180610564S}}
\tablenotetext{d}{The azimuthally averaged gas vertical density profile in
\mi{} is nearly constant to this height, though individual regions show smaller
scale heights and dense clouds.}
\end{deluxetable*}

The other set of simulations are isolated N-body simulations of interactions
between the Milky Way and a Sagittarius-like dwarf galaxy, described briefly
in Section~\ref{ssec:sag_sim}. These include dark matter and stars and are
tailored to existing measurements of the structure of the Milky Way's disk and
of the orbit and properties of the Sagittarius dwarf galaxy. Comparing the
azimuthal midplane variations in the host galaxy of these simulations before
and after the interaction with the Sagittarius-like object gives an idea of
the effect of one minor merger whose properties are relatively well measured.
Azimuthal variations of the mean vertical height of stars has been explicitly
pointed out in a different simulation of a Sagittarius-like encounter by
\citet{2013MNRAS.429..159G}.

\subsection{Description of FIRE Simulations} \label{ssec:cosmozoom}
The FIRE cosmological hydrodynamic simulations
\citep{2014MNRAS.445..581H,2018MNRAS.480..800H} use the zoom-in technique
\citep[e.g.,][]{1993ApJ...412..455K,2014MNRAS.437.1894O} to model the
formation of a small group of galaxies at high resolution in a full
cosmological context. Feedback from supernovae, stellar winds, and radiation
from massive stars is implemented at the scale of star forming regions
following stellar population synthesis models, generating galactic winds
self-consistently \citep{2015MNRAS.454.2691M, 2017MNRAS.470.4698A} while
reproducing many observed galaxy properties, including stellar masses, star
formation histories, metallicities, and morphologies and kinematics of thin
and thick disks \citep{2014MNRAS.445..581H, 2016MNRAS.456.2140M,
2017MNRAS.467.2430M, 2016ApJ...827L..23W, 2018MNRAS.481.4133G,
2018MNRAS.480..800H}.

For this work, we focus on the three Milky Way-mass zoom-ins considered in
\citet{2018arXiv180610564S}, which were simulated as part of the
\textit{Latte} suite and show broad agreement of many of their global
properties with observations of the Milky Way \citep{2016ApJ...827L..23W,
2018MNRAS.481.4133G}. The $\z = 0$ snapshots\footnote{In this work, to avoid
confusion with the vertical height $z$, we refer to cosmological redshift as
$\z$.} of these three simulations, The snapshots of these three simulations at
cosmological redshift $\z = 0$, named \mi{}, \mf{}, and \mm{}, are publicly
available alongside associated mock \textit{Gaia} DR2 catalogues generated
from them.\footnote{\url{http://ananke.hub.yt}}

These simulations contain dark matter particles of mass $\sim35,000\,\Msun$,
gas particles of mass $\sim7000$ to $20,000\,\Msun$, and star particles of
mass $\sim5000$ to $7000\,\Msun$, with the lower end coming from stellar
evolution \citep{2018arXiv180610564S}. Softening lengths for dark matter and
star particles are fixed at $112\,\pc$ and $11.2\,\pc$,
respectively.\footnote{This is $2.8$ times the often-quoted
Plummer-equivalent.} The gas softening length is adaptive, but at $\z=0$ the
median softening length for cold ($T < 100\,\text{K}$) gas particles around
roughly solar positions (with galactocentric cylindrical radii within
$0.5\,\kpc$ of $8.2\,\kpc$ and $\abs{z}<1\,\kpc$) is $53.4$, $57.2$, and
$60.1\,\pc$ for \mi{}, \mf{}, and \mm{}, respectively. These values are
summarized in Table~\ref{tab:scale_height}, along with measurements of the
stellar and gas disk scale heights.

The softening lengths used in the simulations can affect the ability to
resolve the very thinnest planar structures, which in turn can affect how much
the density-based midplane varies as a function of azimuth. The Milky Way's
dense, star-forming gas disk is thought to have a scale height of about
$40\,\pc$, on the order of the cold gas softening length
\citep{2019ApJ...871..145A}. The thin stellar disk has a scale height of about
$300\,\pc$, $\sim30$ times the stellar softening length
\citep{2008ApJ...673..864J}. We therefore expect that resolution effects are
still affecting the scale heights of these components in the simulations,
especially the cold gas. Indeed, the stellar scale heights of the simulated
galaxies are equal to or larger than the Milky Way's while the gas scale
heights are significantly larger (although the proper basis comparison is less
clear in the case of the gas; the quoted value for the Milky Way comes from
studies of high-mass star-forming regions). The midplanes defined by gas and
stars can be tilted with respect to one another as well, precluding extending
the precision of the gas midplane definition to the stellar component.

Cosmological simulations of Milky Way-mass galaxies are not perfect
representations of the true Milky Way in other ways as well, as discussed in
\citet{2018arXiv180610564S}. The failure of cosmological simulations to
exactly reproduce the Milky Way is not necessarily due to limitations of the
numerical model. Candidate Milky Way-like galaxies are chosen solely on their
mass and isolation, for which there are a wide variety of possible galaxies
with qualitatively different properties. For example, the velocity structure
of \mi{} is closer to M31's than the Milky Way's (S. Loebman et al., in
preparation).

However, in this work we are most interested in the global properties of the
potential, and specifically in deviations from axisymmetry. From this
perspective, the simulated galaxies are actually \emph{more} axisymmetric than
we might expect of the Milky Way. While they have prominent spiral arms, none
has as strong a bar as the Milky Way does at present day, and none has a
nearby companion like the Large Magellanic Cloud. One of the three we consider
(\mf) does have an ongoing interaction with a satellite galaxy similar in mass
to Sagittarius, which has punched through the Galactic disk outside the solar
circle, leaving behind some of its stars and inducing warping in the disk.

In this work, we take the galactocentric coordinate system described in
Section 3 of \citet{2018arXiv180610564S} as our fiducial coordinate system for
each galaxy. In short, the center of the galaxy is found iteratively. The
center of mass velocity is then determined by all star particles within
$15\,\kpc$ of this center. The galaxy is then rotated onto a principal axis
frame determined by stars younger than $1\,\Gyr$ inside of the fiducial solar
radius $R_{0} = 8.2\,\kpc$, such that the disk plane is the $x$--$y$ plane.

\subsection{Description of Milky Way-Sagittarius Interaction Simulation}
\label{ssec:sag_sim}
In addition to the cosmological zoom-ins, we will also briefly consider
results from a live N-body simulation of a Sagittarius-like encounter. This
simulation offers us the ability to see how the midplane varies in a more
controlled environment. The simulation is described by
\citet{2018MNRAS.481..286L}, but we briefly summarize the most relevant
details here.

For the Milky Way, the dark halo is modeled as a Hernquist sphere of mass
$10^{12}\,\Msun$ and scale length of $52\,\kpc$ \citep{1990ApJ...356..359H},
the disk is modeled as an exponential disk with a scale radius of $3.5\,\kpc$,
scale height $0.53\,\kpc$, and mass $6\times10^{10}\,\Msun$, and the bulge as
a Hernquist sphere of mass $10^{10}\,\Msun$ and scale length $0.7\,\kpc$. The
Sagittarius dwarf is modeled with two components: a dark matter Hernquist
sphere of mass $8\times10^{10}\,\Msun$ and scale length $8\,\kpc$, and a
stellar component modeled as a Hernquist sphere of mass $6.4\times10^8\,\Msun$
and scale length $0.85\,\kpc$. All components are realized with distributions
of live N-body particles; the Milky Way and Sagittarius are each initialized
to be in equilibrium in isolation.
 
The mass resolution of the simulation is $2.6\times10^4$, $1.2\times10^4$, and
$1.0\times10^4\,\Msun$ for the dark matter, disk, and bulge components,
respectively. For the disk and bulge components, a softening length of
$30\,\pc$ is used whereas for the halo a softening length of $60\,\pc$ is
used. For Sagittarius, the softening length for the dark matter and the stars
is $60$ and $40\,\pc$, respectively.

The fiducial coordinate system for these N-body simulations is the rest frame
of the aligned host galaxy at the beginning of the simulation.

\subsection{The Local Midplane} \label{ssec:local_midplane}
Using the two sets of simulations, we determine the local midplane as a
function of azimuth at the solar circle that an observer might measure if they
were situated in each of these galaxies. Starting from the coordinate system
described in the previous section, which is aligned so that the $z$-coordinate
is approximately perpendicular to the disk plane at the solar circle, we place
our imaginary observer at $z=0$ and a galactocentric cylindrical radius of
$8.2\,\kpc$ and vary the azimuth between $0<\phi<2\pi$. At equally spaced
values of $\phi$ we then compute the median $z$ for stars within a cylinder of
radius $0.5\,\kpc$ and height $1\,\kpc$ perpendicular to the fiducial disk and
centered on it. We choose to use 50 bins in azimuth, sufficiently few that no
cylinder shares stars with its neighbors. We then re-define the new midplane
of the cylinder to be the median $z$, re-select stars, and iterate until the
median $z$ value converges. We find that only $10$ iterations of this
procedure are necessary for convergence. The resulting median $z$ is taken to
be what our observer would measure as the local galactic midplane at each
$\phi$.

\begin{figure*}[htb!]
\begin{center}
\includegraphics[width=\textwidth]{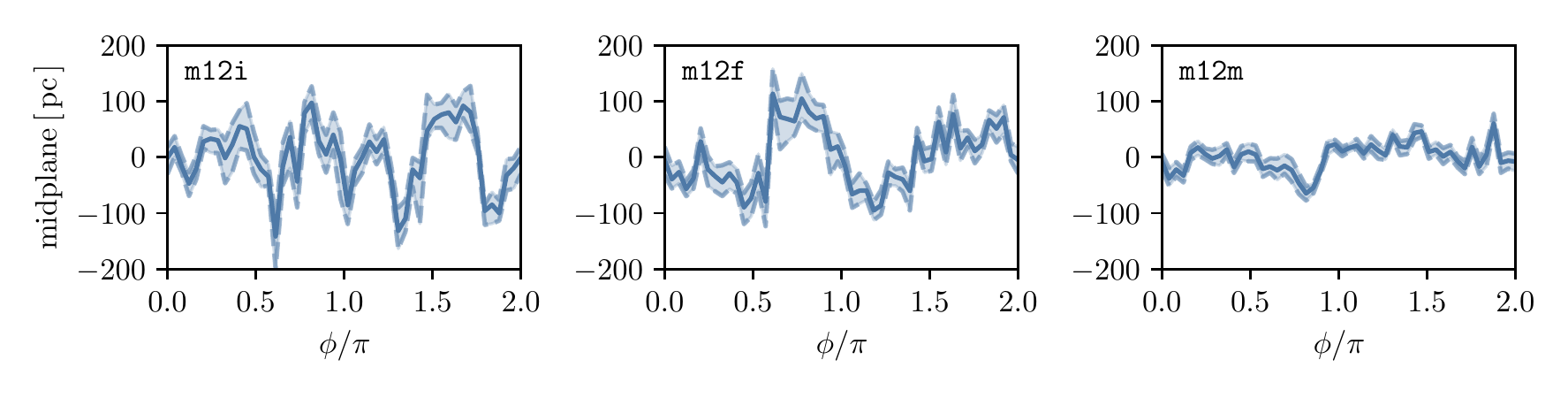}
\end{center}
\caption{The local midplane determined at the fiducial solar circle
($R_0 = 8.2\,\kpc$) for the three FIRE galaxies \mi{}, \mf{}, and \mm{} (left,
center, and right panels) as a function of azimuthal angle, at cosmological
redshift $\z =0$. The local midplane is determined at a position $\phi$ by
taking the median height of all stars within $R=0.5\,\kpc$ and $z=1\,\kpc$ (in
cylindrical coordinates). In order to allow for the possibility that the
fiducial galactocentric coordinate system is incorrect, we subtract the best
fit sine curve from each panel. We then bootstrap resample $1000$ times to
determine $1\,\sigma $ error bars, which we report as dashed lines.}
\label{fig:midplane}
\end{figure*}

\begin{figure*}[htb!]
\begin{center}
\includegraphics[width=342.078286667pt]{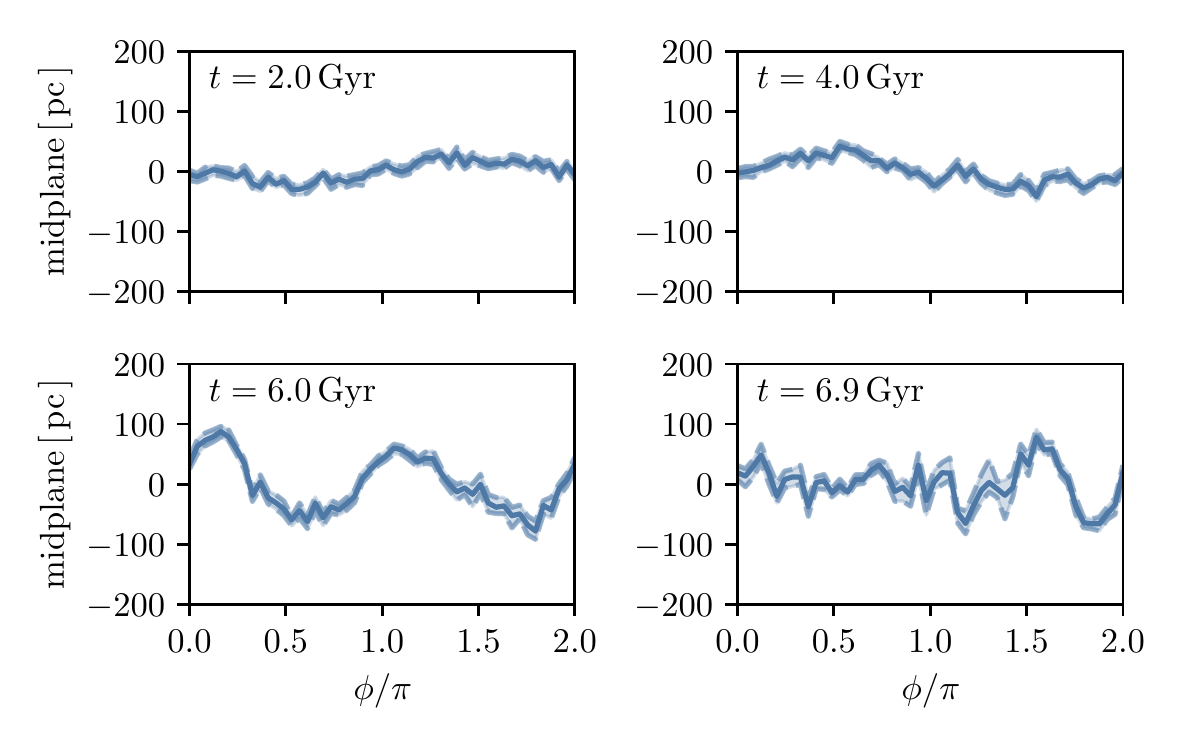}
\end{center}
\caption{The local midplane determined at the fiducial solar circle
($R_0 = 8.2\,\kpc$) for four different time steps from a live N-body
simulation of a Sagittarius encounter with the Milky Way
\citep{2018MNRAS.481..286L}. As before, we have subtracted the best fit sine
curve to account for inaccuracies in the galactocentric coordinate system.
Error bars are calculated as in Figure~\ref{fig:midplane}. The upper panels
show the midplane as a function of azimuth before the first encounter near the
solar circle at $t=2.0\,\Gyr$ and $t=4.0\,\Gyr$, with an encounter happening
close to the solar circle near $t=6.0\,\Gyr$. The fact that the $t=6.9\,\Gyr$
panel, which shows the midplane variation after some relaxation, looks
qualitatively similar to the panels from the FIRE simulations
(Figure~\ref{fig:midplane}) is evidence that midplane variations are
generated, in part, by mergers.}
\label{fig:midplane_chervin}
\end{figure*}

This procedure assumes perfect density estimation, and therefore perfect
corrections for extinction within the cylinder defining the ``solar
neighborhood.'' Imperfect extinction correction is likely to increase the
amplitude of the estimated fluctuations in $z$.

To account for the effect of particle noise, we bootstrap resample stars
within a cylinder of height $2\,\kpc$ and the same radius $1000$ times and
determine the $1\,\sigma$ error bars by repeating the midplane determination
with that reselection.

To allow for potential small inaccuracies in the determination of the original
fiducial coordinate system, we also subtract the best fit curve of the form
\beq \label{eq:fit-curve}
A \sin{\left(\phi + B\right)} + C\text{,}
\eeq 
from the midplane as a function of azimuth to account for an overall tilt of
the midplane (a simplified version of the strategy described in
\citealt{2019ApJ...871..145A}). For simulations (\mi{}, \mf{}, \mm{}), the
best fit values are $A = (-170, 45, 8.8)\,\pc$, $B = (38, -5.0,
1.8)\,\text{deg}$, and $C = (-69, 19, -18)\,\pc$. For the assumed solar radius
of $8.2\,\kpc$, we can approximate the angle offset $\Delta \theta$ for the
$z$-axis from the values of $A$. We find for the same simulations $\Delta
\theta = (1.15, 0.31, 0.062)\,\text{deg}$. These angle offsets are consistent
with the values given in \citet{2018arXiv180610564S} for the difference
between the $z$-axis as defined by the gas and stars.

Figure~\ref{fig:midplane} shows the relative $z$ location of the inferred
midplane an observer would determine as a function of azimuth for each galaxy,
using their local solar neighborhood (the cylinder defined above). The
$1\,\sigma$ error from the bootstrap procedure is shown as the dashed-line
error bars. The middle $90\%$ of midplane values across the solar circle spans
$(190, 160, 84)\,\pc$ for these simulations. In two of the three cases the
midplane therefore varies by more than $\pm 100\,\pc$ depending on the azimuth
along the solar circle; in the third (\mm{}, which has the thinnest ``thin
disk'' of stars, but the largest stellar mass) the variation is closer to~$\pm
50\,\pc$.

We compute the same midplane variation in Figure~\ref{fig:midplane_chervin},
but for four succcessive timesteps of the live N-body simulation of a
Sagittarius encounter \citep{2018MNRAS.481..286L}. Again we have subtracted a
best fit curve of the form given in Equation~(\ref{eq:fit-curve}), with the
values at times
    (2.0, 4.0, 6.0, 6.9)~Gyr being $A$ = (9.5, 2.5, -210, -390)~pc,
    $B$ = (0.074, 0.039, -36, -57)~deg, and $C$ = (2.6,-7.8, -65,
    -53)~pc. The middle $90\%$ of midplane values across the solar
    circle spans (49, 62, 140, 120)~pc. 

These values for the midplane variation are consistent with the azimuthal
midplane variations seen by \citet{2013MNRAS.429..159G}. However, they only
saw significant variations in their Heavy but not their Light Sagittarius
model (virial masses of $10^{11}\,\Msun$ and $\sim3\times10^{10}\,\Msun$,
respectively). The model we used (L2 from \citet{2018MNRAS.481..286L}) has a
virial mass of $6\times10^{10}\,\Msun$, intermediate between their two models.

In the upper panels, we see that the midplane is relatively flat in the inner
galaxy, but additional encounters drive strong midplane variation. In the
lower left panel, we see a strong $m=2$ mode develop, consistent with the
$R=8\,\kpc$ panel of Figure~17 in \citet{2018MNRAS.481..286L} ($m=0$ and $m=1$
modes are stronger, but these are removed in our sine-curve subtraction). The
lower right panel, which shows the galaxy at $t=6.9\,\Gyr$ when some
relaxation has occured, is qualitatively similar to the midplane variations we
saw in the FIRE simulations (Figure~\ref{fig:midplane}), evidence that they
are at least partially driven by mergers.

\subsection{Velocity Variations} \label{ssec:lsr_var}
We also expect that the LSR should vary as a function of azimuth. We perform
this calculation in Appendix~\ref{app:lsr} to estimate the components of the
LSR as a function of azimuth, but performing a best-fit subtraction to correct
for misalignment of the original coordinate system (as in the previous
section) is more involved. Since we find that the variation in the LSR is less
pronounced than for the midplane, and since offsets in velocity only
contribute to second order to $\Delta J_R$ and $\Delta J_z$ when a star is at
maximum amplitude in $R$ or $z$ (where the majority of the orbit is, see
Section~\ref{ssec:epi_action}), we defer this calculation to future work.

\section{Discussion} \label{sec:discussion}
We have used high-resolution simulations to illustrate why we expect the local
midplane defined by stellar density to vary with azimuth by up to $\pm
100\,\pc$ as a natural consequence of the non-axisymmetry of the Galactic disk
at small scales. While this is not in itself surprising or new, we have also
argued that the discrepancy between our local midplane and that of distant
stars introduces a systematic error in the $z$ component when converting from
heliocentric to Galactocentric coordinates. This systematic error introduces a
non-Gaussian error in the vertical action, $J_z$, when starting from the
present-day positions and velocities of stars as measured by, e.g.,
\textit{Gaia}.

These systematic errors are most important for stars on thin disk-like orbits,
where they can be large enough to yield actions representative of orbits in
the thick disk. This effect is entirely due to the extension of a local to a
global coordinate system, and is separate from real diffusion in stellar
integrals of motion caused by interactions with these same deviations from
axisymmetry, such as resonant perturbation by spiral arms or scattering from
molecular clouds \citep{2014RvMP...86....1S}.

\subsection{Estimates of Milky Way Midplane Offsets}

\label{ssec:mw_data_midplane}

Systematic variations in $v_z$ and number density were first noted as
asymmetries in the local velocity distribution towards the North and South
Galactic Caps from the radial velocity surveys of the Sloan Digital Sky
Surveys \citep{2012ApJ...750L..41W} and RAdial Velocity Experiment
\citep{2013MNRAS.436..101W}. Subsequently, \citet{2013ApJ...777L...5C} pointed
out suggestions of an oscillation in average vertical velocities of order
$5\,\kms$ on roughly kiloparsec scales looking toward the Galactic anticenter.

Work by the \textit{Gaia} collaboration confirmed these preliminary results on
the velocity and spatial scales of oscillation with clear spatial maps made
using DR2 data of median $v_z$ over a significant Galactic volume
\citep{2018A&A...616A..11G, 2019arXiv190209569F}, which can be explained with
models of Sagittarius-like encounters
\citep{2013MNRAS.429..159G,2018MNRAS.481..286L,2019MNRAS.485.3134L}. We also
see in the FIRE simulation that the vertical velocity variation as a function
of azimuth is $\sim5\textup{--}10\,\kms$ (Figure~\ref{fig:lsr_variations}),
consistent with these observations.

The vertical frequency of \thin{} and \thick{} are $\sim 0.09\,\Myr^{-1}$ and
$\sim0.06\,\Myr^{-1}$, respectively (Table~\ref{tab:orbits}). By dimensional
analysis, and assuming a vertical velocity variation of
$5\textup{--}10\,\kms$, we therefore expect the midplane offsets to be $\sim
57\textup{--}170\,\pc$. We stress that this is a rough calculation.

Three-dimensional dust maps also offer a view into the expected variation of
the stellar disk, since dust should trace regions of massive star formation.
Figure~9 of \citet{2019MNRAS.483.4277C}, Figure~1 of
\citet{2019arXiv190105971L}, and Figure~2 of \citet{2019arXiv190502734G} all
show that the midplane varies by $\sim10\degree$ at a distance of
$\sim0.75\,\kpc$, corresponding to a physical vertical variation of
$\sim130\,\pc$.

Already we see evidence in the data from velocities and dust maps for midplane
offsets on the order of what we saw in both sets of simulations.

\subsection{Uncertainties in the Solar Position and Velocity}\label{ssec:coord_off}

Uncertainties in measurements of the position and velocity of the Sun relative
to the Galactic center can also contribute to systematic error in the actions,
since converting from heliocentric to Galactocentric coordinates relies on
these measurements. Therefore, errors in their values will induce a systematic
offset in the Galactocentric phase-space position of any observed star.
Considerable effort has been placed on each of these measurements, but
uncertainties remain, and detailed modeling across the disk\,---\,particularly
for dynamically cold stars\,---\,may have to take them into account. Here we
briefly review the current measurements of the four relevant quantities, their
uncertainties, and the implications for the calculation of actions.

\subsubsection{Galactic Center Position}
First, one must define the center of the Galaxy. This is usually taken to be
the location of the central supermassive black hole,
Sagittarius~A\textsuperscript{*} \citep[\sgra{},
e.g.,][]{2004ApJ...616..872R}. From stellar motions near
\sgra{}, the distance from the Sun to \sgra{}, $R_0$, can be
precisely measured \citep{2009ApJ...692.1075G, 2018AA...615L..15G}. A recent
measurement using near-infrared interferometry places $R_0$ at $8.178 \pm
0.035$~kpc \citep{2019arXiv190405721A}, or a $0.4\%$ uncertainty.

However, the location of \sgra{} may not be equivalent to the location of the
dynamical Galactic center, the point in three-dimensional space about which
the stars in the solar neighborhood are orbiting. This assumption, although
sensible and frequently made, has not yet been justified.

If the dynamical Galactic center is offset from \sgra{} by $100\,\pc$, only a
$1.2\%$ difference, then this induces a $\sim15\%$ error in $J_R$ for the
disk-like orbits we considered (see Section~\ref{ssec:quant}). The reason such
a large error in $J_R$ can be generated by a small error in $R_0$ can be
understood from the epicyclic approximation
(Equation~\eqref{eq:Ji_err_mosttime}), which states that $\Delta J_R/J_R =
2\Delta R/A_R$. The fractional error in $J_R$ is related to the error in $R_0$
as a fraction of the \emph{radial amplitude} of the orbit, which is much
smaller than $R_0$ ($\sim1.2\,\kpc$ for \thin{} and \thick{}). This also
implies the very precise $0.4\%$ measurement of $R_0$ still translates to a
$\sim6\%$ uncertainty in $J_R$.

The assumption that the dynamical Galactic center and \sgra{} are colocated is
tested in any construction of a dynamical model where $R_0$ is a free
parameter. For example, \citet{2015ApJ...803...80K} measured $R_0$ while
modeling the dynamics of the stream Palomar 5. Many other dynamical
measurements of $R_0$ have been made (\citealt{2016ARAA..54..529B} summarize
many pre-\textit{Gaia} results), but none have yet achieved a precision
comparable to that of the distance to \sgra{}.

We did not consider in this work the effect of the angular position of the
dynamical Galactic center being offset from \sgra{}.

\subsubsection{Galactic Orientation}
Second, one must define the angular orientation of the Galaxy. This was
defined in 1958 by the IAU subcomission 33b \citep{1960MNRAS.121..123B} by
defining the coordinates of the Galactic center in B1950 coordinates as
(17:42:26.6, -28:55:00) and the North Galactic pole as (12:49:00, +27:24:00).
These two quantities, together with $R_0$, define the orientation of the
Galactic plane. However, there is growing evidence that the stellar midplane
is tilted relative to this coordinate system \citep{2014ApJ...797...53G,
2016ARAA..54..529B}, though not the H~\textsc{ii} midplane
\citep{2019ApJ...871..145A}.

This tilt will contribute a systematic offset in $z$, with the exact magnitude
depending on the position of the observed star. For instance,
\citet{2014ApJ...797...53G} quote a $\sim0.4\degree$ tilt at $3.1\,\kpc$,
corresponding to a vertical height of $\sim22\,\pc$. This corresponds to a
$37\%$ error in $J_z$ for \thin{} and a 5\% error for \thick{}.

\subsubsection{Solar Height}
Third, one must define the Sun's vertical distance from the Galactic midplane,
which can be determined by identifying where the stellar density and
velocities reach a maximum (effectively the median height of all disk stars).
The solar height is usually taken to be $\sim 25\,\pc$
\citep{2001ApJ...553..184C}, with a more recent measurement from \textit{Gaia}
DR2 placing it at $20.8 \pm 0.3\,\pc$ \citep{2019MNRAS.482.1417B}. Another
strategy is to use the cold gas or H~\textsc{ii} regions in the disk to define
the Galactic midplane, leading to slightly different values (by $\sim 5\,\pc$)
for the Sun's relative height \citep[e.g.,][]{2019ApJ...871..145A}. A
pre-\textit{Gaia} review of these measurements is given by
\citet{2016ARAA..54..529B}. The discrepancy between gas-based and
stellar-based determinations of the solar height is small, and thus only
likely to dominate over intrinsic midplane variations on small scales, but
will be relevant for detailed modeling of young and kinematically cold stars.
For instance, it will induce a $\sim10\%$ error in $J_z$ for an orbit with
$z_{\text{max}}\sim100\,\pc$.

\subsubsection{Local Standard of Rest}
Finally, one must define the LSR, or mean velocity of stars near the Sun
relative to the Galactic center (which is defined to have zero velocity), and
the velocity of the Sun relative to the LSR. The radial ($U_{\odot}$) and
vertical ($W_{\odot}$) components are computed by taking the mean motions of
different stellar groups \citep[e.g.,][]{2012MNRAS.427..274S}. The azimuthal
component ($V_{\odot}$) is more difficult to measure, but can be modeled using
the asymmetric drift relation \citep{2008gady.book.....B}. The values of the
components of the LSR are usually taken from \citet{2010MNRAS.403.1829S}.
Their uncertainties should also lead to systematic errors in the actions, as
given in Equations~\eqref{eq:induced_Jphi}--\eqref{eq:induced_Jz}. For
example, the value of the circular velocity is taken to be $\sim 220\,\kms$
\citep[e.g.,][]{2012ApJ...759..131B} with roughly $10\%$ uncertainty. We
expect this to translate to at least a $10\%$ systematic error in $J_{\phi}$.

\subsection{Orbit Integration}\label{ssec:orbit_integrate}
We have mainly been concerned with actions, since they provide a convenient
way to quantify different types of orbits. However, all of our conclusions
also apply to studies that simply rely on orbit integrations, since the two
are equivalent. For instance, computing orbital properties of open or globular
clusters \citep[e.g.,][]{2016A&A...588A.120C, 2018A&A...615A..49C,
2018A&A...616A..12G} should ideally take the midplane variation into account.
Orbit integrations of nearby systems over short time periods
\citep[e.g.,][]{2014MNRAS.445.2169M, 2018A&A...616A..37B} are unlikely to be
impacted. It should also be unimportant for halo applications, e.g., in
modeling of stellar streams \citep[e.g.,][]{2014ApJ...795...95B} or the
substructure potentially responsible for the gap in GD1
\citep{2018arXiv181103631B}.

\section{Conclusions}\label{sec:conclusion}
Determining the orbital properties of stars is important for understanding the
structure and evolution of the Galaxy. Actions have been argued to be
excellent orbit labels. If the Galaxy can be well approximated as axisymmetric
and 6D phase space positions can be measured accurately and precisely enough,
then the computed actions are invariant with orbital phase. However, we have
shown that the fact that the Galactic midplane is not constant across the disk
presents a significant complication to computed actions actually being
invariant. Our main conclusions are:

\begin{itemize}
\item Inaccuracy in the Galactocentric coordinate
system induces orbital phase dependence in the actions calculated from the
observed positions and velocities of stars
(Figures~\ref{fig:cartoon}~and~\ref{fig:one_orbit_wrong_ref}). Since stars'
instantaneous phase-space positions are measured without prior knowledge of
their orbital phases, this results in systematic error in the computed actions
(Figure~\ref{fig:many_orbit_wrong_ref}).

\item Inaccuracy in the midplane location most severely affects computation of
the vertical action $J_z$. A midplane offset of $\sim15\,\pc$ for a typical
thin disk orbit results in a $25\%$ error in $J_z$, and even for a thick disk
orbit a $120\,\pc$ offset will result in the same size error. The fractional
error is significantly less for halo orbits.

\item The distribution of systematic errors in the actions induced by a
coordinate system offset is highly non-Gaussian. The distribution is bimodal
with \emph{neither mode at null}. As a result, error propagation of coordinate
system offsets is complex when considering actions, and is likely to
significantly deform the action-space distribution function.

\item Dynamical modeling across large regions of the disk, over which the
midplane location varies by more than the limits discussed above, is
susceptible to this type of systematic error, since the assumption that our
local Galactic midplane is the global Galactic midplane is not true a priori.
A violation of this assumption (by, e.g., intrinsic midplane variations) leads
to a systematic error in $z$ which generates the large errors in actions
summarized above.

\item 
We show that such midplane variation is likely by measuring the local galactic
midplane along the solar circle in three different high-resolution, zoom-in
simulations of Milky Way mass galaxies from the FIRE collaboration, as well as
a controlled simulation of the interaction of the Milky Way with Sagittarius.
We found that the midplane varies as a function of azimuth at the solar circle
by $60\textup{--}185\,\pc$ in these simulations.

\item Assuming a vertical velocity variation of the Milky Way of
$\sim5\textup{--}10\,\kms$, consistent with recent results from \textit{Gaia}
and our results from the FIRE simulations (Figure~\ref{fig:lsr_variations}),
we estimated that the corresponding midplane offsets are
$\sim60\textup{--}170\,\pc$ by dimensional analysis using the vertical
frequencies of disk-like orbits. This range of values is consistent with the
variations seen in the simulations. Similar offsets are seen in
three-dimensional dust maps.

\item Inaccuracies in the parameters of the currently adopted
Galactocentric coordinate system are likely important for some applications.
In particular, it is imperative to test the assumption that the dynamical
Galactic center is colocated with \sgra{}. We discuss how to do this in
Section~\ref{ssec:coord_off}.

\item This work underlines the importance of combining chemistry and dynamics.
Since chemical tagging \citep{2002ARA&A..40..487F} is not subject to the same
systematic errors discussed in this work, it should be used to confirm
dynamical associations and to offset the effect of these systematic errors on
the action-space distribution function.

\item While in this work we have focused on systematic errors in action
computation, all of our conclusions also extend to studies of stars that
simply rely on orbit integration, since the computation of actions and orbit
integrations are equivalent.

\end{itemize}

Our main point is that the local midplane varies between different points in
the Galaxy, and that this variation can lead to significant systematic errors
in the computation of actions under the assumption of a global axisymmetric
potential. Current observations from \textit{Gaia} should soon permit a
measurement of the real azimuthal dependence of the midplane location. For
some applications, such as those using actions as labels to group stars on
similar orbits, using such a measurement to shift stars to a consistent
midplane height as a function of azimuth before using a global axisymmetric
approximation to the potential may be sufficient, although this ignores the
\emph{dynamical} implications of shifts in the midplane height (which result
from fluctuations in the local density). However, for other applications, such
as the study of action diffusion, a more extensive perturbative approach is
likely needed. We plan to explore the mitigation of these effects in future
work.

\vspace{12pt}

\subsection*{Supplementary Material}
All code used in this work is available at
\url{https://github.com/gusbeane/actions_systematic}.

\acknowledgments
We thank the anonymous referee for providing helpful comments. We would like
to thank Megan Bedell, Robert A. Benjamin, Maria Bergemann, Joss
Bland-Hawthorn, Tobias Buck, Elena D'Onghia, Benoit Famaey, and Adrian
Price-Whelan for helpful discussions. A.B. would like to thank Todd Phillips
for helpful discussions. This work uses data hosted by the Flatiron
Institute's FIRE data hub. The Flatiron Institute is supported by the Simons
Foundation. This project was developed in part at the 2019 Santa Barbara Gaia
Sprint, hosted by the Kavli Institute for Theoretical Physics (KITP) at the
University of California, Santa Barbara. This research was supported in part
at KITP by the Heising-Simons Foundation and the National Science Foundation
(NSF) under Grant No. PHY-1748958. This work used the Extreme Science and
Engineering Discovery Environment (XSEDE), which is supported by NSF Grant No.
OCI-1053575. A.B. was supported in part by the Roy \& Diana Vagelos Program in
the Molecular Life Sciences and the Roy \& Diana Vagelos Challenge Award.
K.V.J.'s contributions were supported in part by the NSF under Grant No.
AST-1715582. M.-M.M.L. was partly supported by the NSF under Grant No.
AST-1815461.

\software{This work made use of the following software: \texttt{astropy}
\citep{astropy:2013,astropy:2018}, \texttt{gala} \citep{gala, gala:zenodo},
\texttt{matplotlib} \citep{Hunter:2007}, \texttt{numpy} \citep{numpy:2011},
\texttt{scipy} \citep{scipy:2001}, and \texttt{tqdm}
\citep{casper_o_da_costa_luis_2019_2800317}.}

\appendix
\section{Orbits} \label{app:orbits}
\begin{figure*}[htb!]
\begin{center}
\includegraphics[width=\textwidth]{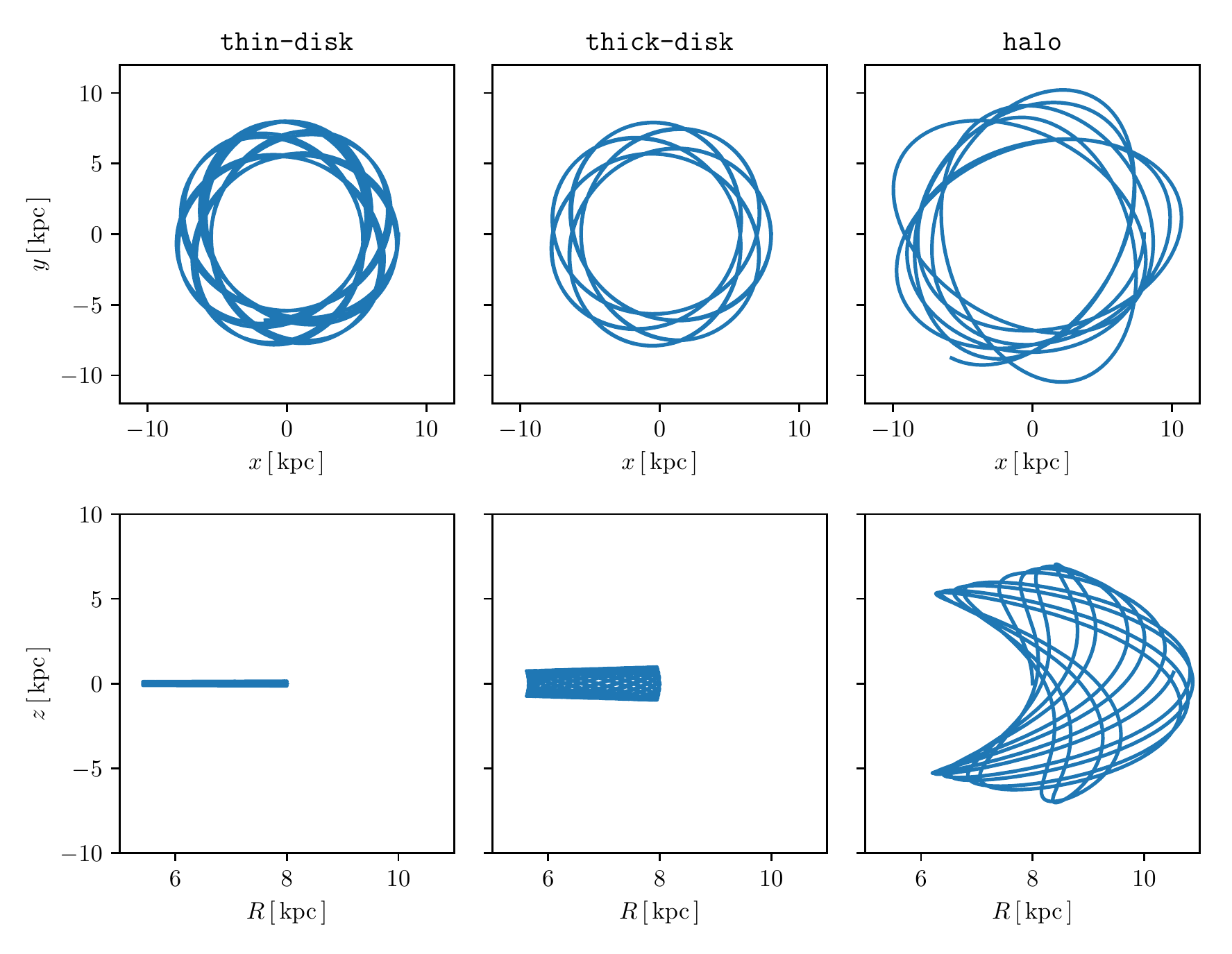}
\end{center}
\caption{The three orbits presented in Table~\ref{tab:orbits} and considered
throughout the work. We plot \thin{}, \thick{}, and \halo{} in the left,
center, and right columns, respectively. The upper row shows a plot of
$x$~vs.~$y$ while the lower row shows $R$~vs.~$z$.}
\label{fig:plot_orbits}
\end{figure*}

We plot the three orbits considered throughout the work
(Table~\ref{tab:orbits}) in Figure~\ref{fig:plot_orbits}.

\section{$\Delta R$-$\Delta x$ Relation} \label{app:deltax}
We considered the effect on actions of an inaccuracy in the distance from the
Sun to the Galactic center, which introduces an offset in the $x$ coordinate,
$\Delta x$, of each star when converting to a Galactocentric coordinate
system. In observations of nearby stars, we have that $\Delta x \sim \Delta
R$. However, for the experiment we performed in Section~\ref{ssec:quant} we
considered observations of a star throughout its entire orbit. Therefore, we
must average $\Delta R$ over the course of the orbit. We derive this relation
now.

An offset $\Delta x$ results in an erroneous radius $R_{\text{err}}$ related by
the formula,
\beq
(x+\Delta x)^2 + y^2 = R_{\text{err}}^2\text{.}
\eeq
Keeping only terms to first order in $\Delta x$, we have that,
\beq
\begin{split}
R_{\text{err}}^2 &= R^2 - 2 R \cos{\phi} \Delta x \\
\implies \Delta R &\equiv \abs{R_{\text{err}} - R}
        = \abs{\cos{\phi}} \Delta x\text{.}
\end{split}
\eeq
Averaging over the circle, we therefore have that,
\beq
\avg{\Delta R} = \frac{2}{\pi} \Delta x\text{.}
\eeq

\section{$J_R$ and $J_{\phi}$ Distributions} \label{app:hist}

\begin{figure*}
\begin{center}
\includegraphics[width=7.10000594991in]{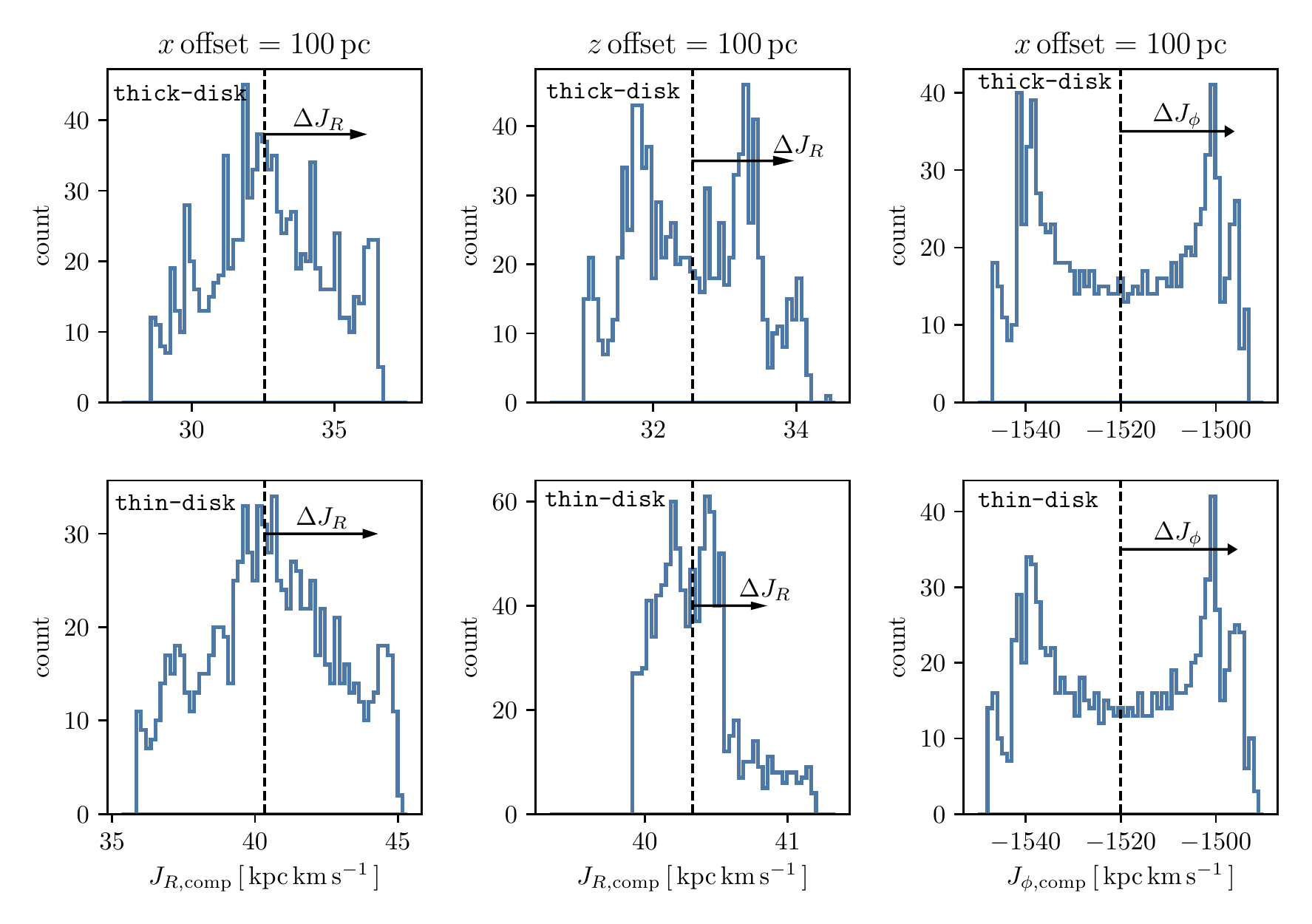}
\end{center}
\caption{A histogram of the values computed for $J_R$ and $J_{\phi}$ for
\thick{} (upper panels) and \thin{} (lower panels). For $J_R$ we assume an
$x$ offset (left) and $z$ offset (center) of $100\,\pc$, while for $J_{\phi}$
we consider only an $x$ offset (right). In each panel the true value is given
by a vertical dashed line. The induced error distribution in $J_R$ for an $x$
offset more closely resembles a Gaussian centered on the null value, but not
for the other two offsets considered.}
\label{fig:Jphi_JR_hist}
\end{figure*}

In Figure~\ref{fig:Jphi_JR_hist} we plot the distribution of $J_R$ as a
function of orbital phase induced by an offset in $x$ and $z$ and the
distribution of $J_{\phi}$ for an offset in $x$. We plot the distributions for
\thick{} (upper panels) and \thin{} (lower panels). We find that the $J_R$
distribution induced by an offset in $x$ more closely resembles a Gaussian
distribution, while the $J_R$ distribution induced by an offset in $z$ and the
$J_{\phi}$ distribution induced by an offset in $x$ are both similar to the
$J_z$ distribution induced by an offset in $z$ (see Figure~\ref{fig:Jz_hist}).

\section{LSR Variations} \label{app:lsr}
\begin{figure*}[htb!]
\begin{center}
\includegraphics[width=\textwidth]{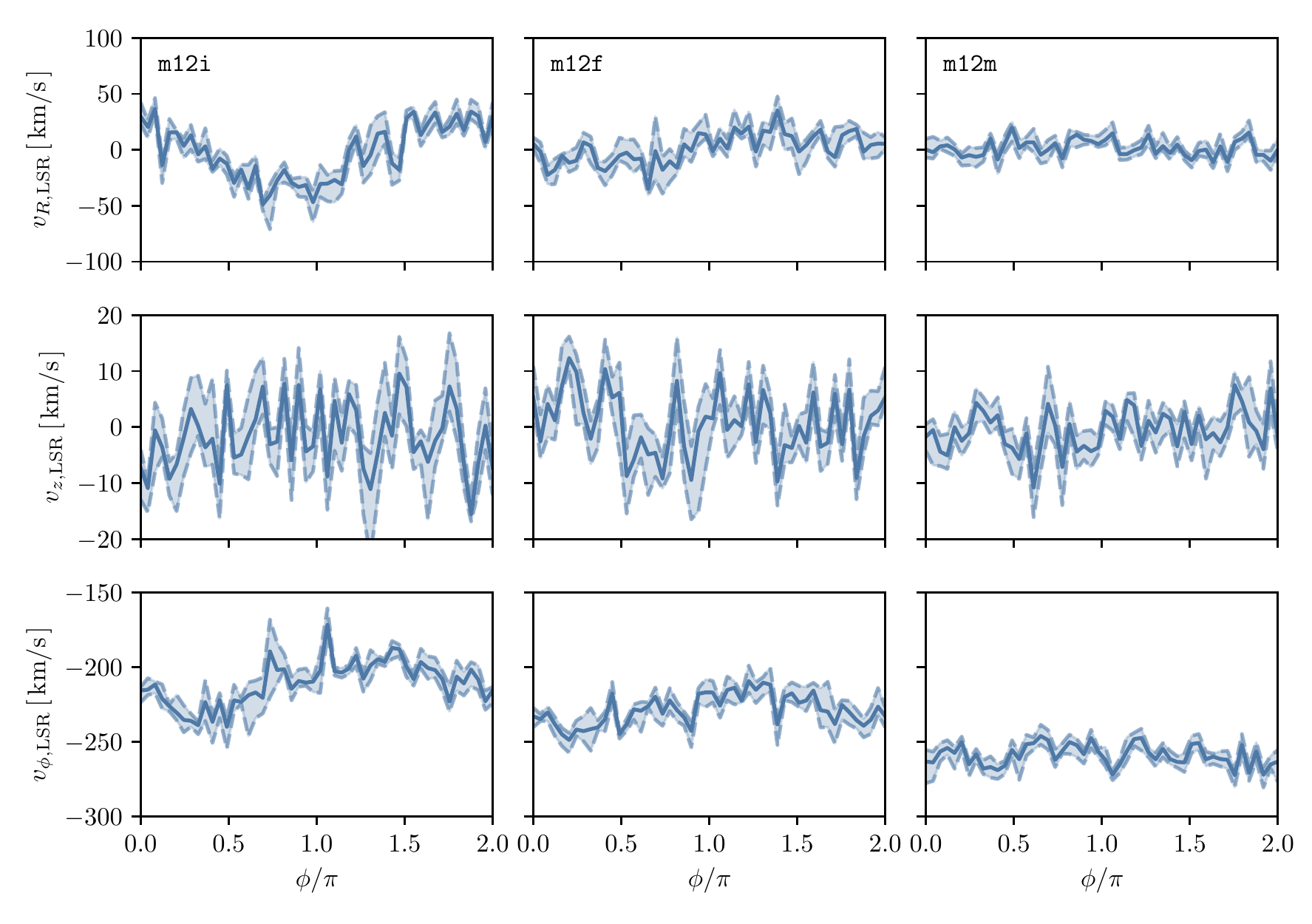}
\end{center}
\caption{The LSR as a function of azimuth at the fiducial solar circle ($R_0 =
8.2\,\kpc$). No best-fit subtraction is performed here as we did in the case
of the midplane (Section~\ref{ssec:local_midplane}). Variations in $v_z$ are
on the order of $\sim5\textup{--}10\,\kms$.}
\label{fig:lsr_variations}
\end{figure*}

We consider the variations of the LSR as a function of azimuth at the fiducial
solar circle ($R_{0} = 8.2\,\kpc$) in Figure~\ref{fig:lsr_variations}. At each
azimuth, $\phi$, we take the median velocity in cylindrical coordinates of all
stars within $200\,\pc$ of the position, following
\citet{2018arXiv180610564S}. No best-fit subtraction was performed as in
Figure~\ref{fig:midplane}.

\bibliography{references}

\end{document}